\documentclass{article}
\title{A complete simulation
  framework for stone
  degradation on 3D real
  geometries}

\author{$^\S$ Silvia Preda, ${}^\dagger$ Gabriella Bretti, $^\ddagger$ Francesco Freddi,\\ $^\ddagger$ Bruno Nazarena and $^\S$ Matteo Semplice\\ \\ \small
\small$^\ddagger$Department of Engineering and Architecture, University of Parma, Italy \\
 \small$^\dagger\!\!$ National Research Council of Italy, Institute for Applied Mathematics ``M. Picone'', Rome, Italy\\
 \small$^\S$ Dept. of Science and High Technology, University of Insubria, Como, Italy \\}

\usepackage{amsmath,amsfonts}
\usepackage{graphicx}        %
\graphicspath{{img/}}
\usepackage{url}

\usepackage{booktabs}
\newcommand{\R}{\mathbb{R}}

\newcommand{\DT}{\ensuremath{\mathrm{\Delta}t}}
\newcommand{\DX}{\ensuremath{\mathrm{\Delta}x}}
\newcommand{\DY}{\ensuremath{\mathrm{\Delta}y}}
\newcommand{\DZ}{\ensuremath{\mathrm{\Delta}z}}

\newcommand{\intd}{\ensuremath{\mathrm{d}}}

\newcommand{\pder}[2]{{#1}_{#2}}
\newcommand{\PDER}[2]{\frac{\partial #1}{\partial #2}}

\usepackage{nicefrac}

\newcommand{\poros}{\rho}
\newcommand{\internal}{\mathcal{I}}
\newcommand{\ghosts}{\mathcal{G}}

\usepackage[textsize=footnotesize,backgroundcolor=yellow!70,bordercolor=orange]{todonotes}

\renewcommand{\todo}[2][]{\relax}

\begin{document}

\date{}
\maketitle

\begin{center}\sl
    Dedicated to the memory of our colleague and friend Maurizio Ceseri, that unfortunately passed away in 2024 during the development of this project.
\end{center}

\begin{abstract}
We present a complete workflow for predicting stone degradation phenomena, such as marble sulfation, in works of art. The main challenge is to accurately acquire the geometry of the artwork and then use it to perform simulations based on a mathematical model of the degradation process, typically formulated as a system of partial differential equations (PDEs).

To address this, we generate a point cloud of the object surface using photogrammetric techniques and subsequently post-process it to obtain a level-set description of the three-dimensional geometry. This representation is then incorporated into the numerical discretization of the PDE system. Combined with suitable time-stepping and preconditioning strategies, the resulting framework enables the prediction of degradation evolution, such as the growth of gypsum crust thickness on marble, under different scenarios.
\end{abstract}

\section{Introduction}
\label{sec:intro}
Conservation of cultural heritage is of crucial importance for the world cultural, social and economic development. 
Regrettably, part of this precious heritage is now in serious danger, because of accelerating degradation processes of monumental stones related to weathering and climate changes. Degradation phenomena are often due to the overlapping of environmental, chemical factors and mechanical actions, caused by the synergistic action of atmospheric agents and pollutants. Such phenomena may determine an irreversible weakening of the mechanical strength and an increased vulnerability to chemical aggressions of stones.

\begin{figure}
    \centering
    \includegraphics[width=0.75\linewidth]{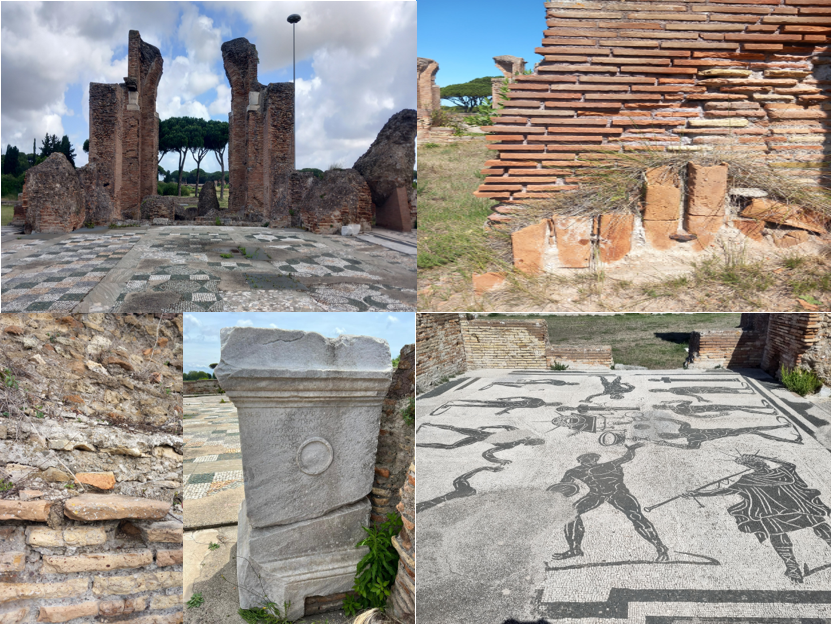}
    \caption{Some pictures took from the excavations of Porta Marina spas, in the area of the excavations of the Archaeological Park of Ostia Antica.}
    \label{fig:ostia}
\end{figure}

In this work we present some applications, including a case study on a marble altar situated at the Porta Marina spas in the Archaeological Park of Ostia Antica (https://www.ostiaantica.beniculturali.it/en/home/), near Rome. The park covers a total area of over 150 hectares, with about 33 to 40 hectares containing excavated ancient structures and buildings from Ancient Roman civilization, like the Porta Marina spas, the ancient port, the necropolis,
and the synagogue. The buildings are made of heterogeneous materials: travertine, bricks, marble, terracotta, limestones, etc., see Fig. \ref{fig:ostia}, and they can be affected by different degradation mechanisms brought by atmospheric pollution, rising damp and salty air, due to the proximity to the sea. 

Mathematical models of increasing complexity have been developed by cultural heritage scientists to assist with various aspects of the predictive maintenance of works of art, and the employment of models based on partial differential equations (PDEs) has been advocated for the next generation of models with regulatory powers \cite{reviewSQB:18}. These models are particularly suited to study the current state and predict the future development of a specific work of art, as they operate on a real three-dimensional representation of the object under consideration.
On the other hand, detailed information on the piece of art must be acquired and inserted into mathematical framework.

Degradation mechanisms are described in mathematical models through a set of differential equations, typically involving diffusion terms which describe the penetration of some exogenous substance into the pores of the stone, and some reaction terms that describe the chemical reactions causing the degradation. Often, such degradation changes the material properties of the stone (e.g. its porosity) and thus influences the diffusion terms \cite{carb, salt}. 
Among the most harmful factors there is pollution in the air. 
The most widespread and damaging air pollutants are nitric oxide, carbon dioxide and sulphur oxide. Here we focus on chemical damage caused by sulphur oxide $SO_2$ on building heritage materials. The damaging effects of $SO_2$ were studied in \cite{ABST:12:kinetic, ADN:sulfation, giavarini2003nonlinear, GIAVARINI200814} also taking into account swelling phenomena \cite{CFN:08:swelling}, surface rugosity and mechanical damage \cite{BONETTI2023103886}.
In its simplest form, a model describing reactive flows in porous media will include two equations: a diffusion-reaction equation for the variable representing the gas concentration in the pores, and a reaction one for the affected material.
The simplified model of marble sulfation presented in \cite{ADN:sulfation} belongs to this class and it will be employed for the numerical tests in this paper. It will be however clear that our framework can be applied to other degradation mechanisms and to more complex mathematical models.

Several numerical methods are available in the literature for the numerical approximation of reaction-diffusion time-dependent PDEs, but one of the main challenges in the present application is to acquire the three-dimensional geometry of the work of art and employ it in the numerical computations. To this end we have elected to use simple finite-differences methods on Cartesian grids, coupled with ghost-point methods and a levelset description of the computational domain.

We consider a background Cartesian lattice with points $(x_i,y_j,z_k)\in\R^3$ with $x_{i+1}-x_i=\DX$, $y_{j+1}-y_j=\DY$, and $z_{k+1}-z_k=\DZ$, on which the PDE is discretized with finite-difference approximations. The object geometry is encoded in the model as a so-called ``levelset function'', i.e. a function  $\varphi:\R^3\to\R$
such that the computational domain, corresponding to the work of art, coincides with the set $\Omega=\{\vec{x}\in\R^3: \varphi(\vec{x})\leq 0\}$ and its boundary is
$\partial\Omega=\{\vec{x}\in\R^3: \varphi(\vec{x})= 0\}$.
An example of a levelset function is the so-called ``signed distance function'' that is defined to be $\varphi_{\text{sdf}}(\vec{x})=-\mathrm{dist}(\vec{x},\partial\Omega)$ if $\vec{x}$ is inside the object, and $\varphi_{\text{sdf}}(\vec{x})=\mathrm{dist}(\vec{x},\partial\Omega)$ otherwise.
This approach has been successfully adopted in various fields where complex or evolving domains are to be treated~\cite{osher1999level,osher2005level,OshSet:1988}. In \cite{CSS:monum,cdss:mach19} these techniques have already been applied to the field of cultural heritage, in two space dimensions.

The task of computing the levelset function for a specific work of art is itself non trivial, but it can be tackled as follows. First a point cloud is acquired in-situ, providing the coordinates of a set of points belonging to the external surface of the object. This can be achieved for example by 3D laser scanning or photogrammetric methods \todo{espandere e check citazioni} \cite{remondino2011,ReEl:2006}.
Then, a mathematical model evolves a surface with some prescribed tension and shrinks it until it is attached to the points in the cloud. This computation can be performed directly on the levelset function $\phi$ following \cite{Zhao:2000,Kosa2017,PreSe:25,PreSe:26:adaptive}. The stiffness of the surface can be used as a user-controlled parameter to balance the roughness of the surface and the adherence to the point cloud, and it is often useful to counteract the noise in the acquired dataset.

Finally, model parameters such as the porosity, the reaction speeds, the diffusion coefficients, etc, have to be gathered from the literature or acquired with laboratory experiments on material samples. Then, numerical simulations of the degradation process can be run and examined. In case of uncertainties in the parameters, multiple runs in different scenarios can be run, in order to assess which feature or parameter of the degradation under study is the more relevant for its control.

The rest of the paper is organized as follows.
In \S\ref{sec:pcloud} we describe the acquisition and generation of the point cloud for a real object;
in \S\ref{sec:levelset} we describe our method to compute the levelset function from a point cloud;
in \S\ref{sec:PDE} introduce the PDE model and its numerical treatment. In \S\ref{sec:casestudy} we present an application of the entire work-flow to a case study, before drawing some conclusions and present possible future extensions in \S\ref{sec:concl}.

\section{Acquiring the point-cloud}\label{sec:pcloud}

Three-dimensional (3D) digitization techniques are widely employed across several domains, including engineering, architecture, structural analysis, and industrial applications, for the geometric documentation and quantitative assessment of physical objects. These techniques can be broadly classified into range-based methods, such as Terrestrial Laser Scanning (TLS) and structured-light systems, and image-based approaches, namely photogrammetry. Range-based systems directly acquire 3D coordinates by measuring distances between the sensor and the object surface, typically providing dense and metrically reliable point clouds with limited dependence on surface texture. In contrast, photogrammetric approaches reconstruct object geometry from overlapping images using Structure-from-Motion (SfM) for image orientation and Multi-View Stereo (MVS) algorithms for dense reconstruction, enabling the generation of detailed 3D models enriched with high-resolution texture information \cite{remondino2011}.

Both approaches can achieve high levels of geometric accuracy, while exhibiting different and, in many cases, complementary characteristics. Laser scanning is generally characterized by a high degree of automation and standardized acquisition and processing workflows, requiring limited operator intervention and ensuring consistent data quality. Photogrammetry, on the other hand, offers greater flexibility and scalability, lower operational costs, and the ability to simultaneously capture geometric and radiometric information. However, photogrammetric results may be influenced by image quality, illumination conditions, and acquisition geometry, and typically require the definition of scale and reference system through ground control or measured constraints. In contrast, laser-based techniques may be less effective on reflective or translucent materials and generally provide limited radiometric detail. For these reasons, photogrammetry and laser scanning should not be regarded solely as alternative approaches, but rather as complementary techniques that can be effectively combined depending on the application \cite{fassi2013,nicolae2014,farella2022}.

In the present case, considering the relatively small size of the investigated object (approximately $0.60 \times 0.60 \times 1.20$~m) and the need for a rapid and portable acquisition setup, a close-range photogrammetric approach was selected. This solution allows efficient data acquisition using standard photographic equipment while achieving sub-millimetric ground sampling distances and simultaneously capturing surface texture.

The survey was performed using a Sony ILCE-6400 mirrorless camera, 
equipped with a 16~mm focal length lens (E PZ 16--50~mm, fixed at 16~mm), acquiring images at a resolution of $6000 \times 4000$ pixels. A total of 124 images were collected in approximately 30 minutes following a convergent acquisition geometry, consisting of three circular image strips at different heights around the object, with an average camera-to-object distance of approximately 1.15~m. This configuration ensured a high level of image overlap and robust geometric redundancy. The nominal Ground Sampling Distance (GSD) was approximately 0.3~mm/pixel, allowing the detection and reconstruction of fine geometric details on the object surface.

Image processing was carried out using Agisoft Metashape Professional, following a standard SfM-MVS workflow. Image orientation was performed at high quality without downsampling, resulting in 85,374 tie points and a mean reprojection error of approximately 0.8 pixels. Camera self-calibration parameters (including focal length, principal point, and lens distortion coefficients) were simultaneously estimated during bundle adjustment. Subsequently, dense image matching produced a dense point cloud of approximately 8.5 million points, while a polygonal mesh model consisting of over 1 million faces was also generated.

A local reference system was defined, and the model was scaled using measured distances between identifiable features, ensuring a consistent metric reference suitable for comparative and numerical analyses.

To facilitate downstream applications, the dense point cloud was further processed and resampled to generate datasets at different spatial resolutions (1~mm, 2~mm, and 4~mm point spacing), enabling a trade-off between geometric detail and computational efficiency. The final mesh model was cleaned and optimized prior to export for structural modeling purposes.

\begin{figure}[htbp]
    \centering
    \includegraphics[width=0.85\textwidth]{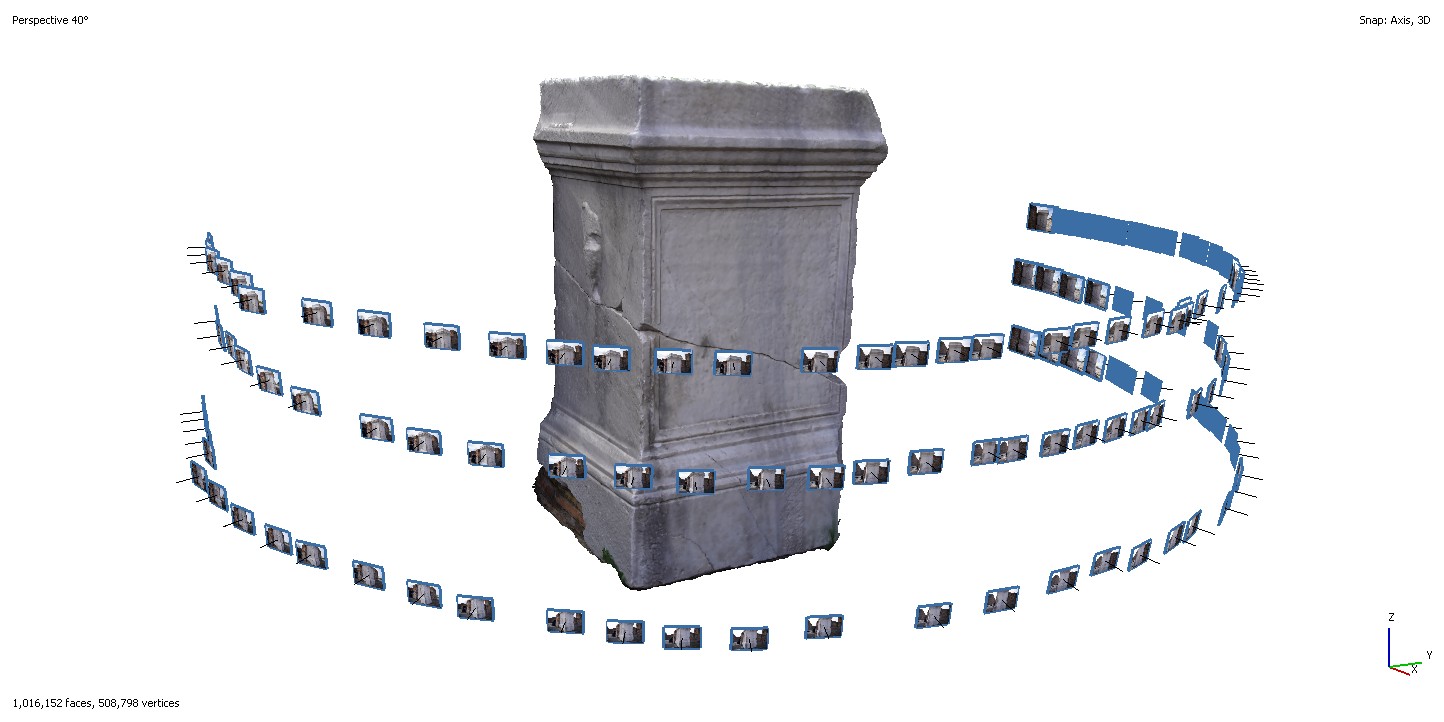}
    \caption{Photogrammetric block geometry with camera positions and the resulting reconstructed mesh model.}
    \label{fig:block_geometry}
\end{figure}

\begin{figure}[htbp]
    \centering
    \includegraphics[width=1.05\textwidth]{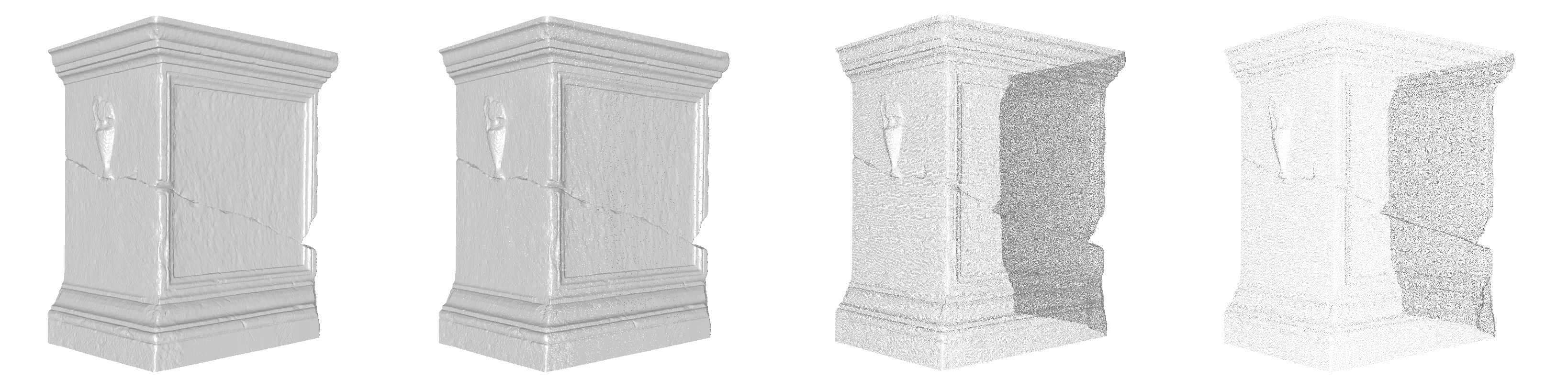}
    \caption{Comparison of point clouds at different spatial resolutions: full-resolution dataset and resampled versions with 1~mm, 2~mm, and 4~mm point spacing.}
    \label{fig:pointcloud_resolutions}
\end{figure}

Figure~\ref{fig:block_geometry} illustrates the image block geometry and the reconstructed model, while Figure~\ref{fig:pointcloud_resolutions} presents examples of the derived point clouds at different spatial resolutions.

\section{Computing the levelset from a point cloud}
\label{sec:levelset}
\newcommand{\pcloud}{\mathcal{S}}

\begin{figure}
    \centering
    \includegraphics[width=0.49\linewidth]{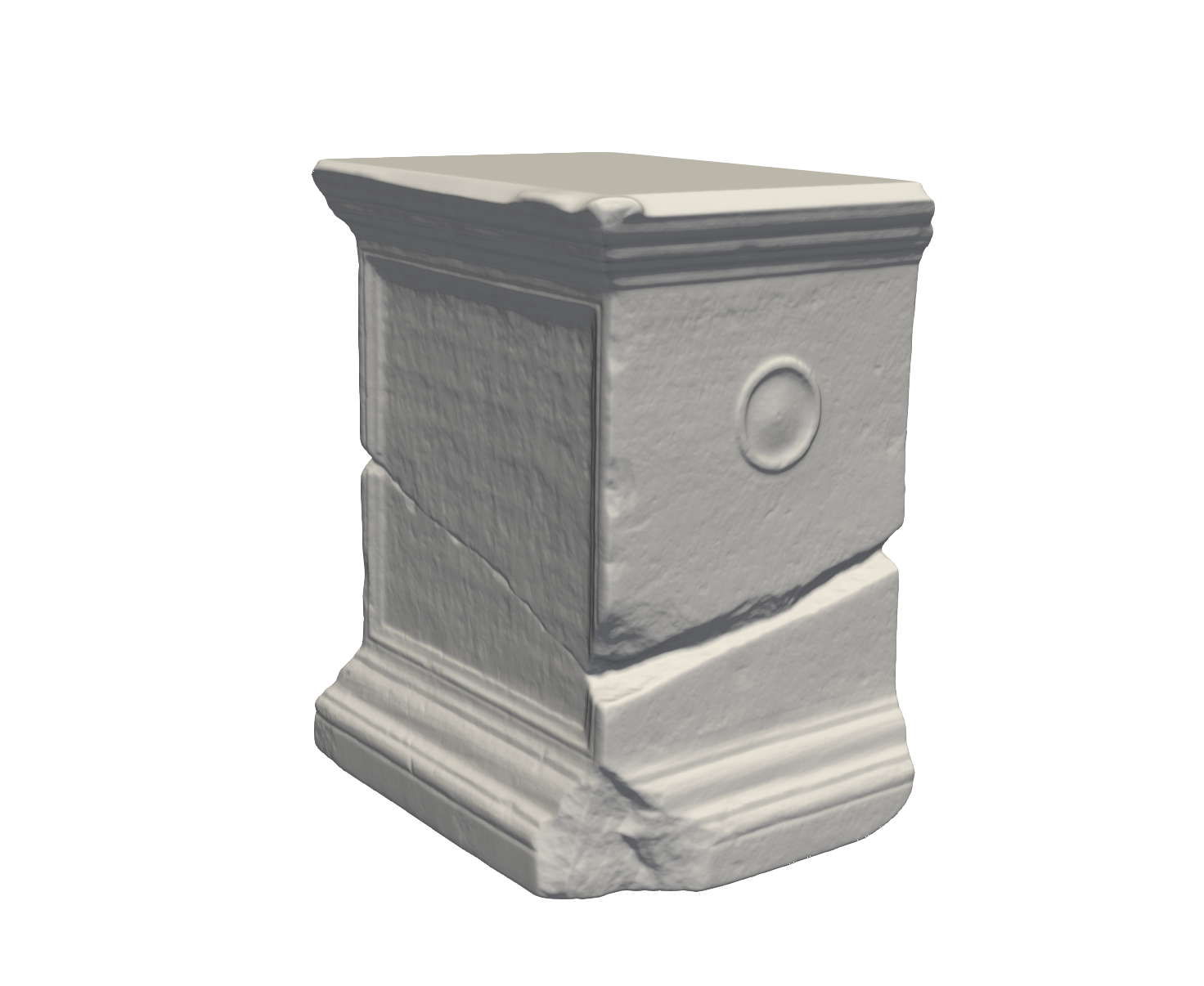}
    \includegraphics[width=0.49
    \linewidth]{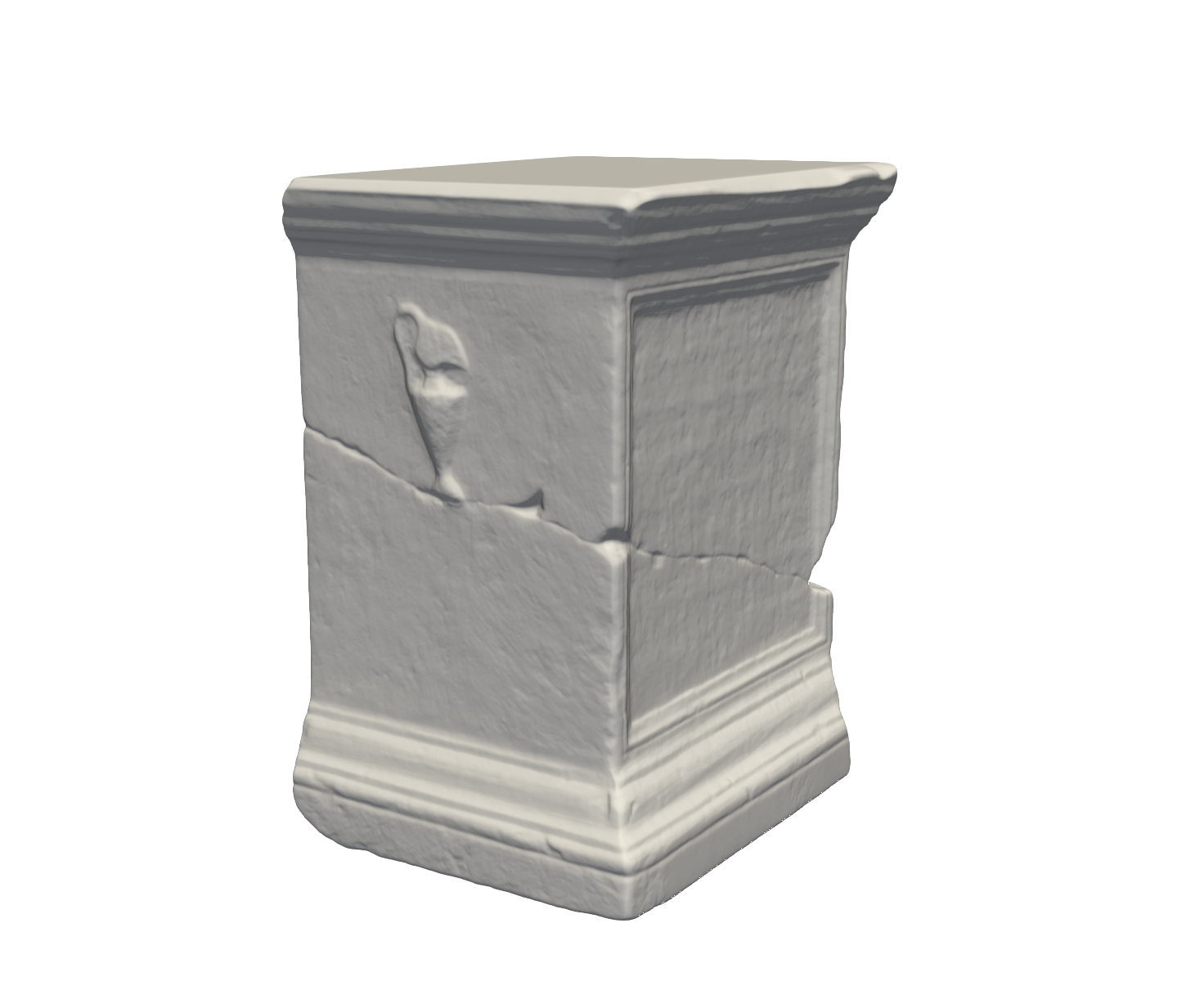}
    \caption{Altar's levelset reconstruction obtained from a point cloud with $1$~mm point spacing. The discretization step $\DX$ is uniform and approximately equal to $1.20$~mm.}
    \label{fig:levelset_altare}
\end{figure}

\begin{figure}
    \centering
    \includegraphics[width=0.4\linewidth]{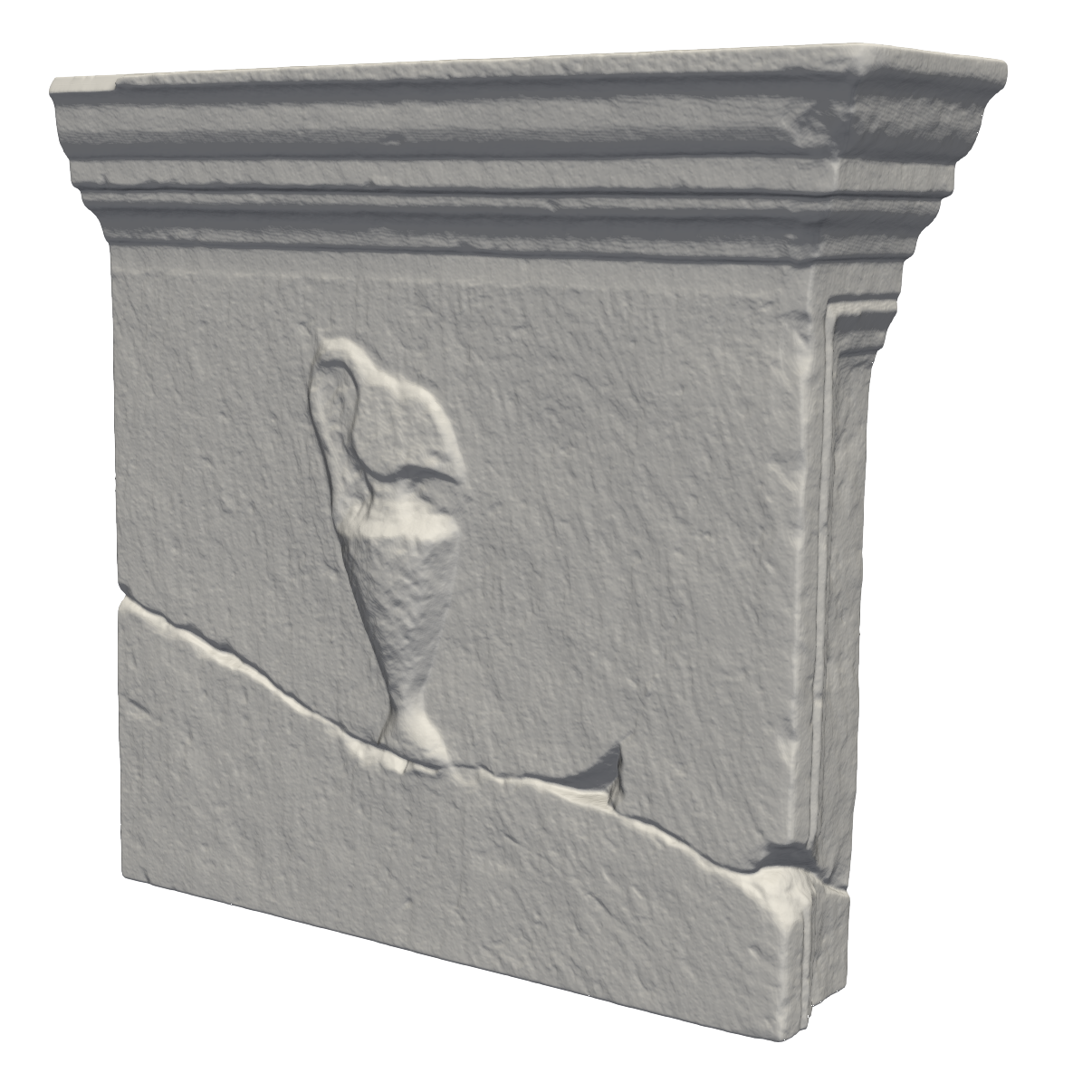}
    \includegraphics[width=0.48\linewidth]{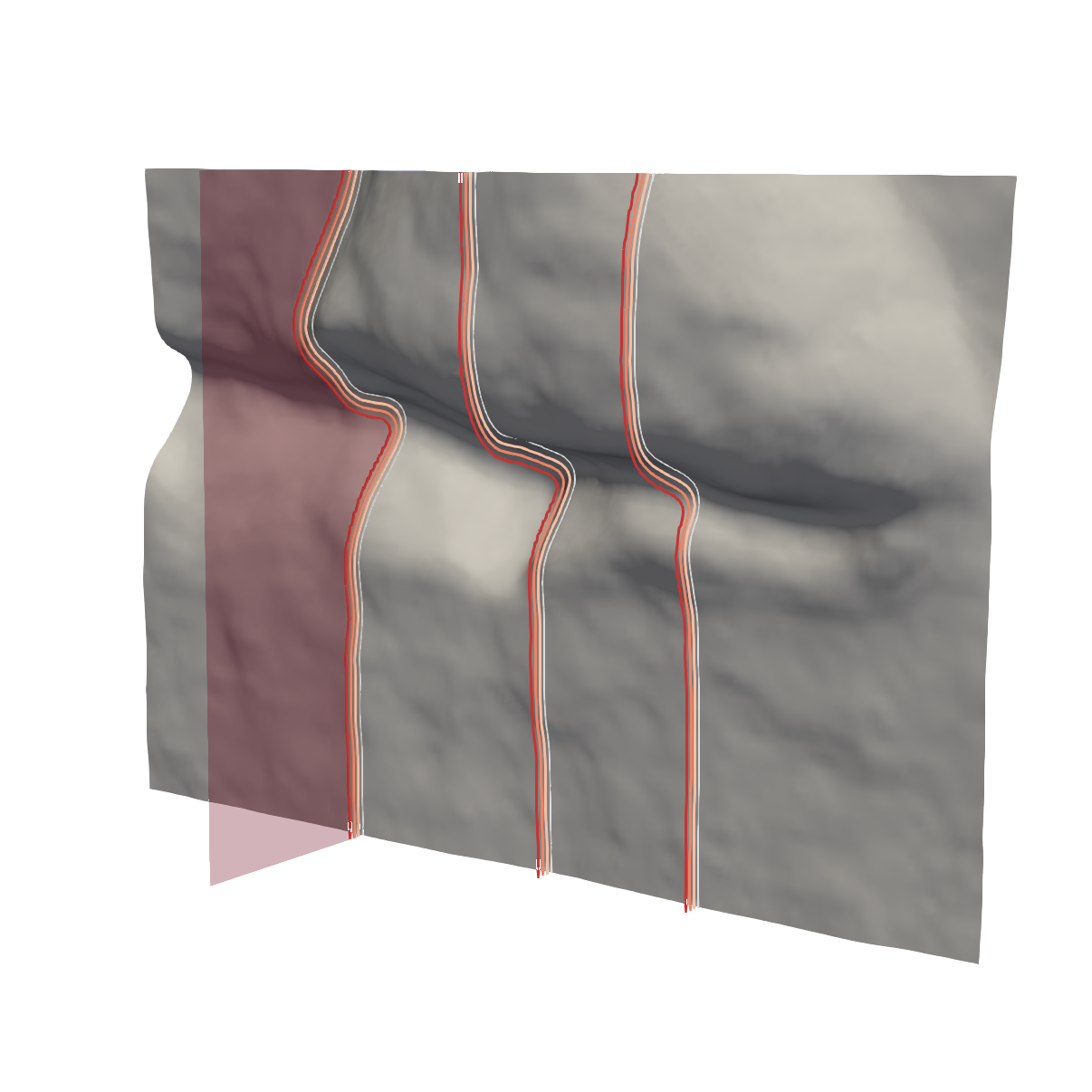}
    \caption{Left panel: Altar's levelset description obtained from a portion of the point cloud related to the amphora on one side. The final levelset is defined on a Cartesian uniform grid of size $1229\times350\times1119$. Right panel: some isocountours of the levelset function. The point cloud resolution is approximately equal to $1$~mm, whereas the discretization step is $\DX \approx 0.53$~mm.}
    \label{fig:levelset_anfora}
\end{figure}

\begin{figure}
    \centering
    \includegraphics[width=0.4\linewidth]{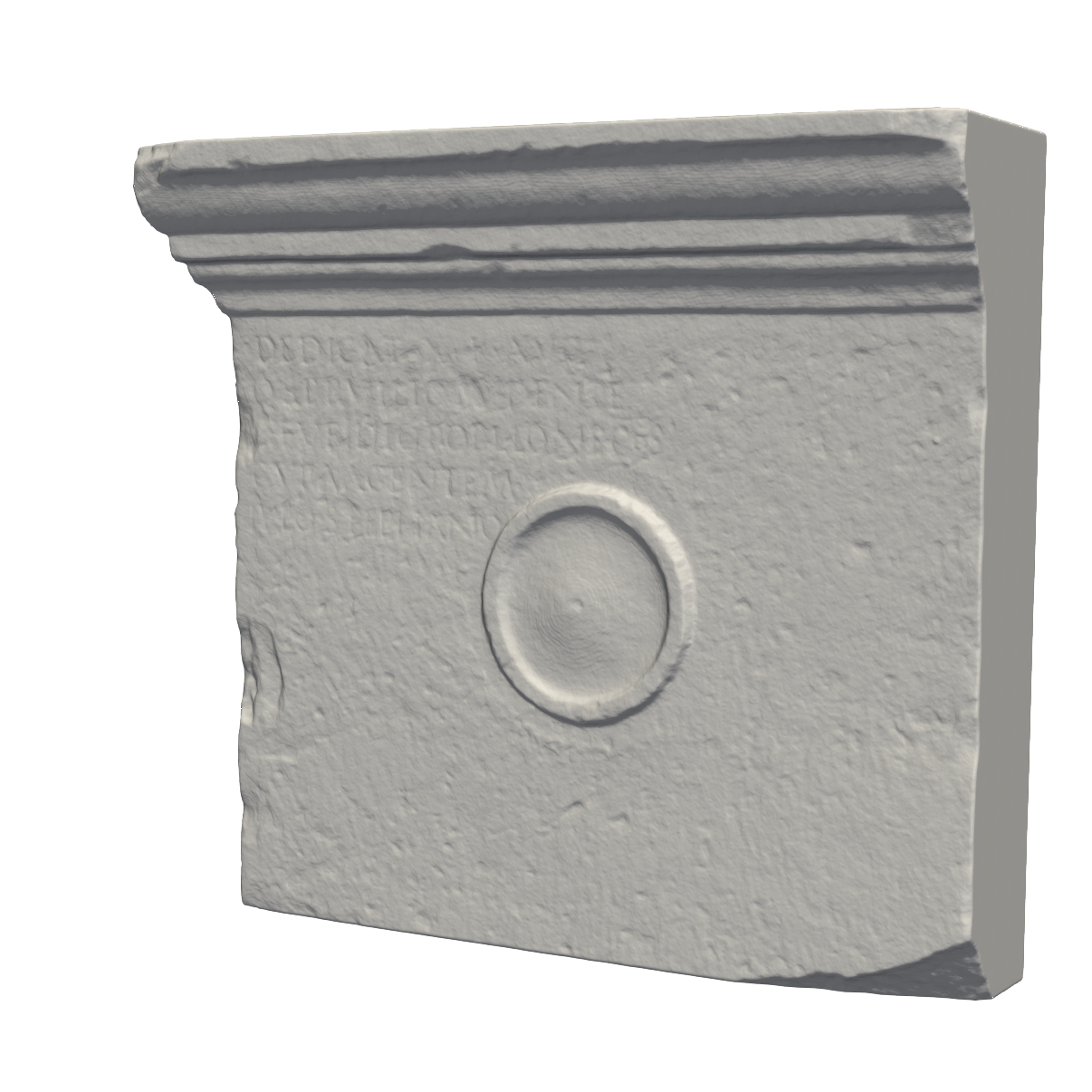}
    \includegraphics[width=0.48\linewidth]{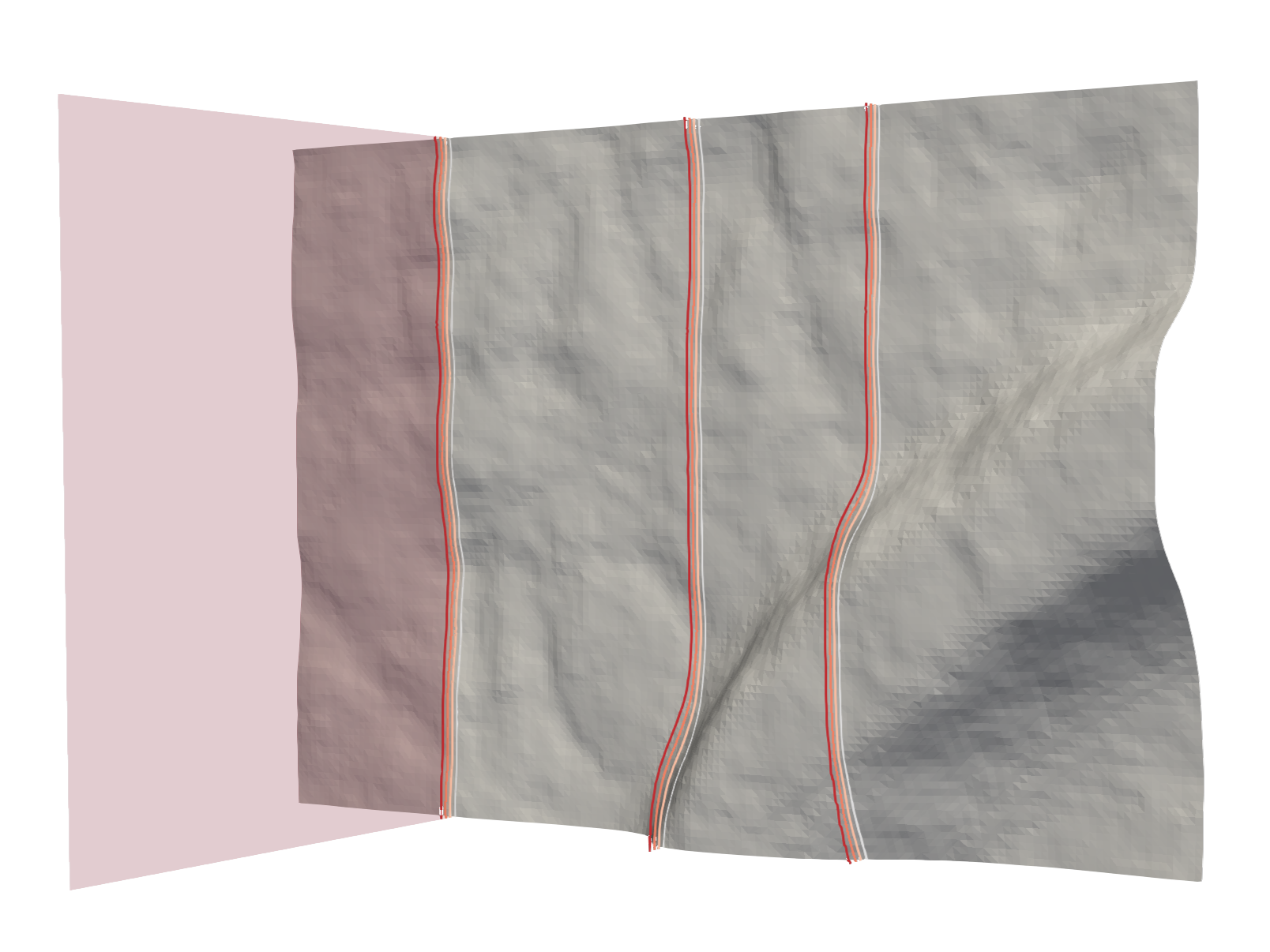}
    \caption{Left panel: Altar's levelset description obtained from a portion of the point cloud related to the inscriptions on one side. The final levelset is defined on a Cartesian uniform grid of size $1229\times350\times1119$. Right panel: a some isocountours of the levelset function. The point cloud resolution is approximately equal to $1$~mm, whereas the discretization step is $\DX \approx 0.52$~mm.}
    \label{fig:levelset_scritte}
\end{figure}

In this section we describe how, from a set of points $\pcloud=\{\vec{q}_1,\ldots,\vec{q}_N\}$ in a bounded region of $\R^3$, we compute a levelset function $\varphi:\R^3\to\R$ that  vanishes on $\pcloud$.
We start from an initial implicit function $\varphi^0(\vec{x})$ that encloses all the data 
and evolve it by shrinking it towards the points in $\pcloud$. In practice $\varphi^0(\vec{x})$ is represented by its values on a fixed background grid, and it is evolved by numerically approximating the levelset equation
\begin{equation}\label{eq:levelset:pde}
\small
\begin{split}
    \pder{\varphi}{t}(\vec{x},t) 
    &= \Bigg[  \frac{d(\vec{x})}{E_p(\varphi)}   \Bigg]^{p-1} \Bigg( \nabla d(\vec{x})\cdot \nabla \varphi(\vec{x},t) + \frac{\mu}{p} d(\vec{x}) \nabla\cdot\left( \frac{\nabla \varphi(\vec{x},t)}{|\nabla \varphi(\vec{x},t)|} \right)|\nabla \varphi(\vec{x},t)| \Bigg),\\
    \varphi(\vec{x},0) &= \varphi^0(\vec{x}),
\end{split}
\end{equation}
until a steady-state is reached. The equation above translates, in the levelset framework, the minimization of the surface energy functional
\begin{equation}\label{eq:energy}
  E_p(\Gamma) = \Big( \int_{\Gamma} d^p(\vec{x})\, \intd s \Big) ^{1/p}, \quad 1 \leq p \leq \infty,    
\end{equation}
where $\Gamma$ is a closed surface of co-dimension one in $\R^n$, corresponding to $\varphi(\vec{x})=0$, and $d(\vec{x}) = \min_{\vec{q} \in \pcloud}|\vec{x}-\vec{q} |$ denotes the Euclidean distance of $\vec{x}\in\R^3$ from the set $\pcloud$.
At steady-state, the surface has reached an equilibrium between the term attracting it towards $\pcloud$ and the curvature term which discourages sharp edges and features; this balance can be controlled by the user through the parameter $\mu\geq0$.

To perform the computations we employ a numerical scheme based on a semi-Lagrangian discretization of both the transport and the curvature term, which is stable for a timestep $\DT$ of the same order as the spatial discretization steps $\DX$, $\DY$, $\DZ$. Given the size of the typical point clouds to be treated, suitable strategies for the parallelization of the code, for concentrating the computational effort in the vicinity of the zero level set, and for the reinitialization of the levelset function $\varphi$, must be employed.
The scheme is described in details in \cite{PreSe:25} for a uniform background Cartesian mesh and has been extended to locally adapted grids in \cite{PreSe:26:adaptive}. 

For the real case presented in this work, the result of the levelset reconstruction phase on a uniform Cartesian lattice are shown in Fig.~\ref{fig:levelset_altare},~\ref{fig:levelset_anfora} and~\ref{fig:levelset_scritte}. The point cloud with $1$~mm resolution has been employed to compute the final levelset, and by selecting some portions of the dataset, we have been able to highlight specific details of the altar, such as the amphora on one side and, on the other, a raised circular feature and engraved inscriptions.

\section{3D numerical simulations of marble sulfation}
\label{sec:PDE}

For the sake of illustrating the full simulation framework for stone degradation of real objects, in this work we employ a very simple model of damage, namely the marble sulfation model of \cite{ADN:sulfation}. In the model the degradation is described by the simplified reaction
\begin{equation*}\label{eq:REAC}
\mathrm{CaCO_3} + \mathrm{SO_2} +\frac12\mathrm{O_2} +2\mathrm{H_2O}
\longrightarrow \mathrm{CaSO_4}\cdot2\mathrm{H_2O} + \mathrm{CO_2}
\end{equation*}
between a sulfur compound present in the pores and the calcium carbonate in the marble, which is turned into gypsum.

We introduce two main variables:
$c(t,\vec{x})$, representing the concentration of marble,
and $s(t,\vec{x})$, representing the concentration of the pollutant,
both defined for $t>0$ and $\vec{x}\in\Omega$,
where $\Omega=\{\vec{x}\in\R^3: \varphi(\vec{x})\leq0\}$ is the computational domain represented by the levelset obtained as in the previous sections.

To take into account that the porosity locally changes while the marble is turned into gypsum, a variable porosity
$
\poros(c(\vec{x},t)) = \alpha \, c(\vec{x},t) + \beta
$  is introduced, where $\alpha$ and $\beta$ are parameters to be fixed as $\alpha=\rho_m-\rho_g$ and $\beta=\rho_g$, in terms of the porosities of the pristine marble ($\rho_m$) and of the gypsum ($\rho_g$) obtained as degradation process.

The model is expressed by the PDE
\begin{equation}\label{eq:PDE}
\left\{
\begin{array}{rcll}
\displaystyle \frac{\partial \left( \poros(c)s \right)}{\partial t} &=& \displaystyle -\frac{a}{m_c} \poros(c) \, s \, c+ d\, \nabla \cdot (\poros(c) \nabla s) & \mbox{ in } \Omega \times [0,T],\\
\displaystyle \frac{\partial c}{\partial t} &=& \displaystyle -\frac{a}{m_s} \poros (c) \, s\, c & \mbox{ in } \Omega \times [0,T]. \end{array}
\right.
\end{equation}
The first terms on the right-hand-side of each PDE represent the reaction, which consumes both $\mathrm{CaCO_3}$ and $\mathrm{SO_2}$, indicated respectively by $c$ and $s$, at speed $a$, being $m_c$ and $m_s$ the masses of the two reactants.
The last term of the first equation of \eqref{eq:PDE} represents the diffusion of the pollutant gas in the stone domain, modelled by a Fickian diffusion process with coefficient $\poros(c)$ and further regulated by the constant $d$.
Of course $a$ and $d$ should be set in accordance to environmental conditions, at least temperature and humidity.

\begin{figure}
    \centering
    \includegraphics[width=0.75\linewidth]{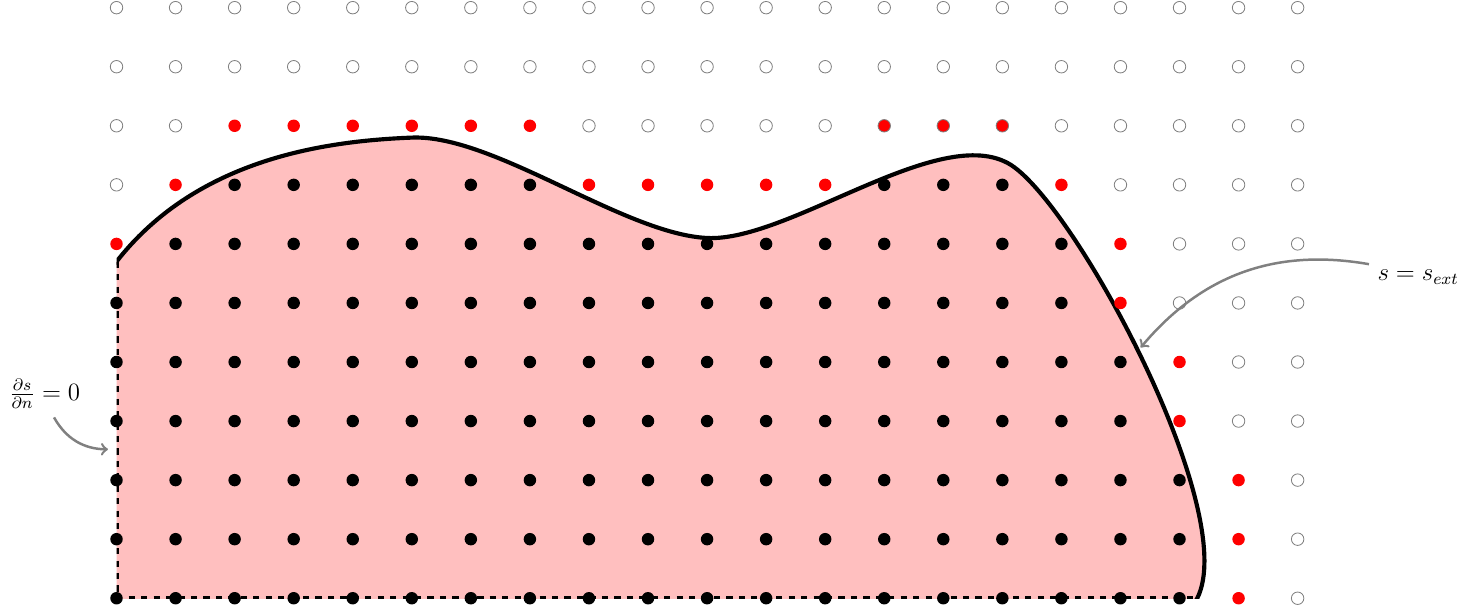}
    \caption{Domain (pink shaded area), internal grid points (black solid dots), ghost points (red solid dots) and inactive grid points (cicles). The Dirichlet (solid line) and Neumann (dashed line) boundaries are also indicated.}
    \label{fig:domain}
\end{figure}

The initial condition for a simulation would typically represent a pristine marble object with no sulfur dioxide inside by considering $c(\vec{x},0)=c_0=1$ and $s(\vec{x},0)=0$ for all $\vec{x}\in\Omega$. For $t>0$, the boundary condition $s(\vec{\gamma},t)=s_{\text{ext}>0}$ for all $\vec{\gamma}\in\partial\Omega$ will trigger the penetration of $\mathrm{SO_2}$ from the surrounding environment through the outer surface of the work of art, towards the interior of the domain. We point out that $s_{\text{ext}}$ could be taken dependent on time and space to represent the variable concentration of air pollutant in the environment during the existence of the work of art.

The task of the numerical model is to compute the time evolution of $s(\vec{x},t)$ and $c(\vec{x},t)$ inside the domain $\Omega=\{\vec{x}\in\R^3:\varphi(\vec{x})\leq0\}$ and apply the boundary condition $\left.s\right|_{\partial\Omega}=s_{\text{ext}}$. The actual computational domain is thus determined by the levelset function $\varphi$ that can be computed with the method described in the previous section.

In Figure~\ref{fig:domain} we illustrate a computational domain $\Omega$ (pink area), immersed in a rectangular  Cartesian grid. The outer boundary of $\Omega$, i.e. the levelset $\{\varphi(\vec{x})=0\}$ is indicated with a solid black line: on this curve, the Dirichlet boundary condition $s=s_{\text{ext}}$ is imposed. If the object is not fully contained in the grid, one has $\{\varphi(\vec{x})<0\}$ along portions of the boundary of the computational grid. This is indicated by dashed lines in the picture and there an homogeneous Neumann boundary condition will be applied.

The background Cartesian mesh is composed by $N_x\times N_y \times N_z$ points, with coordinates
$(x_i,y_j,z_k)\in\R^3$ such that $x_{i+1}-x_i=\DX$, $y_{j+1}-y_j=\DY$, and $z_{k+1}-z_k=\DZ$. In what follows we will denote the values of a generic function $u(\vec{x})$ on the grid points as $u_{ijk}=u(x_i,y_j,z_k)$.

The $N_xN_yN_z$ nodes are divided in three sets (see Fig.~\ref{fig:domain}): the \emph{internal points} $\internal$ (if $\varphi_{ijk})<0$), the \emph{ghost points} $\ghosts$ (if $\varphi_{ijk} \geq 0$ and at least one of the six neighbour points $\left\{ (x_{i\pm1},y_{j},z_{k}), (x_{i},y_{j\pm1},z_{k}), (x_{i},y_{j},z_{k\pm1}) \right\}$ is inside $\Omega$) and the \emph{inactive points} (all the other grid points). These grid points are indicated in black solid dots, red solid dots and black circles in the picture.

In view of the very high reaction constants to be employed, the time discretization is achieved by operator splitting: the reaction terms are approximated by an exponential integrator, and the diffusion term by an implicit finite-difference discretization. This choice avoids the complications of the nonlinear solver employed in \cite{CSS:monum} and more easily guarantees the positivity of the computed concentrations.

The initial condition $c^0_{ijk}=1$, representing pristine marble, and $s^0_{ijk}=0$, representing no pollutant inside the stone, is set on all internal points. Then for each timestep, the approximations
$c^{n+1}_{ijk}$ and $s^{n+1}_{ijk}$ of the variables at internal points are computed from
$c^n_{ijk}$ and $s^n_{ijk}$ in three steps: (a) the reaction terms are evolved with the exponential Rosenbrock-Euler method, (b) a small linear system is solved to set the ghost point values for $s$ on the grid nodes in $\mathcal{G}$ and (c) a large linear system is solved to evolve in time the parabolic equation for the variable $s$.

The exponential Rosenbrock-Euler method \cite{Hochbruck:exp:2010} consists in evolving the ODE system $u'=f(u)$ with the formula $u^{n+1}=u^n + \Psi_1(\DT J_f) f(u^n)$, where $J_f$ is the Jacobian of $f$ evaluated at $u^n$ and $\phi_1$ is the matrix function $\Psi_1(z)=(e^z-1)/z$. In our case, the ODE system to be evolved for every grid point is
\begin{equation}\label{eq:ODE}
  \begin{cases}
  \hat s' = -A_s \hat s c \\
  c' = -A_c \hat s c
\end{cases}
\;,
\end{equation}
where we have introduced the auxiliary variable $\hat s =\rho(c)s$, set $A_s=a/m_c$ and $A_c=a/m_s$, .
The Jacobian matrix is thus $\left[\begin{smallmatrix} -A_s c & -A_s \hat s \\ -A_c c & -A_c \hat s \end{smallmatrix} \right]$, which has eigenvalues $0$, with eigenvector $\left(\begin{smallmatrix} ... \\ ... \hat s \end{smallmatrix} \right)$, and $\ldots$, with eigenvector $\left(\begin{smallmatrix} ... \\ ... \hat s \end{smallmatrix} \right)$. The matrix function is thus readily computed by the explicit diagonalization of $J_f$ and thus ${\hat s}^{n+1}$ and $c^{n+1}$ are computed by
\[
  \begin{pmatrix} \hat s \\ c \end{pmatrix} ^{n+1}
  =
  \begin{pmatrix} \hat s \\ c \end{pmatrix} ^{n}
  +
  \Psi_1(\DT J_f)
  \begin{pmatrix} -A_s \hat s c \\ -A_c \hat s c \end{pmatrix}^n.
\]
Finally, the evolution can be rewritten in terms of $s$ and $c$, by first computing $c^{n+1}$ and then recalling that ${\hat s}^{n+1}=s^{n+1}\rho(c^{n+1})$; one obtains
\begin{equation} \label{eq:RosEul}
\begin{aligned}
  c^{n+1}_{ijk} &= c^{n}_{ijk} \frac{A_s c^{n}_{ijk} + A_c s^{n}_{ijk} \rho(c^{n}_{ijk}) e^{-\DT \delta}}{\delta}
  \\
  s^{*}_{ijk} &= s^{n}_{ijk} \rho(c^{n}_{ijk}) \frac{A_c s^{n}_{ijk} \rho(c^{n}_{ijk}) + A_c c^{n}_{ijk} e^{-\DT \delta}}{\delta \rho(c^{n+1}_{ijk})}
\end{aligned}
\end{equation}
where $\delta=A_c s^{n}_{ijk}\rho(c^{n}_{ijk})+A_s c^{n}_{ijk}$ is the determinant of the $J_f$ matrix.
From equation \eqref{eq:RosEul} it can be appreciated that the computed values for $c^{n+1}_{ijk}$ and $s^{*}_{ijk}$ will remain non-negative even for very large reaction constants $A_s, A_c$.

Formula \eqref{eq:RosEul} sets the new values of all grid points in $\mathcal{I}$ for the $c$ variable at time $t^{n+1}=t^n+\DT$, but the other variable has to be further evolved to take into account the diffusion term. To this end, we apply the Implicit Euler method by solving the linear system
\begin{equation} \label{eq:ie:s}
  s^{n+1}_{ijk} + \DT \mathcal{L}_{ijk}[s] = s^{*}_{ijk}, \text{ for } (i,j,k)\in\internal,
\end{equation}
where $\mathcal{L}_{ijk}[s]$ is a finite-difference discretization of the Laplacian operator on the variable $s$ at the $(i,j,k)$ grid point.

We choose the standard form of the finite-difference Laplacian, namely
\begin{equation} \label{eq:lapl:s}
\mathcal{L}_{ijk}[s]
=
\sum_{P\in\mathcal{N}_{i,j,k}} \frac{\rho(c^{*}_{i,j,k})+\rho(c^*_P)}{2} \frac{s^*_P-s^*_{i,j,k}}{\delta_P}
\end{equation}
where $\mathcal{N}_{i,j,k}=\{(i\pm1,j,k),(i,j\pm1,k),(i,j,k\pm1)\}$ is the set of first neighbours of the grid point $(i,j,k)$ and $\delta_P$ is the distance of $P$ from the central point (i.e. $\DX$, $\DY$ or $\DZ$).

However, the summation in \eqref{eq:lapl:s} may involve points $P\not\in\mathcal{I}$. In this case, since $c^*_P$ is not defined when $P\in\ghosts$, in this case we approximate $\rho(c^*_P)$ with $\rho(c^{*}_{i,j,k})$.
Also $s^*_P$ is not defined at ghost points and the boundary condition $\left.s\right|_{\partial \Omega}=s_{ext}$ must be applied.
We employ a ghost point extrapolation technique that was already successfully adopted in the context of elliptic~\cite{CocoRusso:Elliptic, coco2018second, coco2012second, CCDR:LinearElasticity, coco2016hydro} and hyperbolic~\cite{chertock2018second, CocoRusso:Hyp2012} equations.
To this end, for each $G\in\ghosts$ we construct a stencil $\mathcal{S}_G$ of first neighbours of $G$ that are either internal or ghost points. This is easily achieved by inspecting the signs of $\left.\PDER{\varphi}{x}\right|_{i,j,k}\approx\frac{\varphi_{i+1,j,k}-\varphi_{i-1,j,k}}{2\DX}$ and of the other directional derivatives of the levelset function: in fact, the gradient of the levelset points from the ghost towards the interior of the domain.
A boundary point $B_G\in\partial\Omega$ is also associated to the ghost point $G$ by rootfinding $\varphi(G+\lambda\nabla\varphi|_G)=0$.
Imposing that the linear interpolation of the grid function $s$ on the stencil coincides with $s_{ext}$ at $B_G$ yields the linear equation
\begin{equation} \label{eq:ghost:s}
  \sum_{Q\in\mathcal{S}_G} \alpha_Q s^*_Q = s_{ext}.
\end{equation}
The coefficients $\alpha_Q$ can be computed by the following procedure during the setup phase of the simulation. Choose a basis for linear polynomials in three variables (e.g. $\psi_0=1$, $\psi_1=x-x_G$, $\psi_2=y-y_G$, $\psi_3=z-z_G$) and construct the Vandermonde matrix for $\mathcal{S}_G$ with entries $V_{a,b}=\psi_b(Q_a)$ for $Q_a\in\mathcal{S}_G$. Then the linear (least-square) interpolant of the data in $\mathcal{S}_G$ is given by $\sum_{c=0}^3 \alpha_c \psi_c(\vec{x})$ where the coefficients $\vec{\alpha}$ are computed by multiplying the Moore-Penrose pseuso-inverse $V^{\dagger}$ into the vector $\vec{r}_G$ of the $s^*$ values at the points in $\mathcal{S}_G$. The evaluation of this interpolant at $B_G$ is then $\sum_{c=0}^3 \psi_c(B_G) V^{\dagger}_{cd} (\vec{r}_G)_d$, so that the coefficients $\alpha_Q$ are given by the row vector $\vec{\psi}(B_G)V^{\dagger}$.

Summarizing, $s^{n+1}$ is thus computed by first solving the system of the equations \eqref{eq:ghost:s} for all ghost points and then solving the system \eqref{eq:ie:s} for all internal points. The first is a system of size $|\ghosts|$, which is proportional to the area of the boundary of $\Omega$, while the second is much bigger, having size proportional to the volume of $\Omega$.

The numerical method has been implemented with the help of the PETSc libraries \cite{petsc-web-page} for the domain decomposition of the background Cartesian grid and the solution of the linear systems. For these latter, the smaller system \eqref{eq:ghost:s} is solved by GMRES with ILU preconditioner with zero levels of fill, while the larger symmetric system \eqref{eq:ie:s} is tackled by the Conjugate Gradient method with an Algebraic MultiGrid solver.

\section{A case study}\label{sec:casestudy}

Some parameters of the model can be fixed by literature data. In particular we set
$m_c=100.09$ and $m_s=64.06$ from the molecular weights of the $CaCO_3$ and the gypsum molecule, as in \cite{ADN:sulfation}.
The porosity of marble used in works of art is typically below $1\%$ (see e.g. \cite{porosity}) and we take here the value $\rho_m=0.36\%$, while we set to $\rho_g=50\%$ the porosity of gypsum.

\subsection{Model sensitivity}\label{ssec:sensitivity}
The remaining parameters $a$, $d$ and $s_{ext}$ can be rescaled and only their relative size matters. Typical values of $SO_2$ concentration in air ranges from $15ppb$ in very clean environment to $1ppm$ in very polluted ones. These very low values would however pose problems in the numerical solvers. The $s$ variable, in fact, takes values in the range $[0,s_{ext}]$, while the $c$ variable, which has meaning of the percentage of pristine marble remaining, takes values in the entire range $[0,1]$. The strong separation of the two ranges by several orders of magnitude would cause a strong depletion of the available floating point accuracy, leaving very few significant digits in the computed results.

\begin{figure}
    \centering

    \begin{tikzpicture}
    \fill[cyan!30] (-1,-.5) rectangle (0,.5);
    \fill[pink] (0,-.5) rectangle (10,.5);
    \fill[black] (0,-.1) rectangle (8,.1);
    \draw[->] (-.5,0) -- (10.5,0) node[below]{$x$};

    \node at (0.,0.) [pin={[pin edge={<-,thick}, pin distance=10mm]90:{$s(0,t)=s_{\tiny{ext}}$}}] {};
    \node at (8.,0.) [pin={[pin edge={<-,thick}, pin distance=10mm]90:{$\frac{\partial s}{\partial x}(0.5,t)=0$}}] {};
    \end{tikzpicture}

    \caption{Illustration of the one-dimensional computational domain for the sensitivity test.}
    \label{fig:domain:1d}
\end{figure}
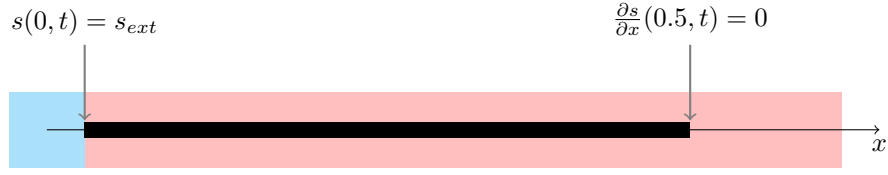

We have thus chosen to rescale $s$ by a reference external $SO_2$ concentration, thus setting $s_{ext}=1$. All the other constants should thus be rescaled accordingly, but their physical values is not easily assessed, since both the diffusion constant $d$ and the reaction rate are dependent on environmental parameters. We thus begin applying the one-dimensional discretization to a very fine grid in order to illustrate the model dependence on the parameters. We consider a one-dimensional domain of $0.5mm$ discretized with $\DX=0.5\mu m$. The initial conditions are $c(0,x)=1$ corresponding to pure marble and $s(0,x)=0$ corresponding to no $SO_2$ present inside. The pollutant is introduced in the system by applying a Dirichlet boundary condition $s(t,0)=s_{ext}$ at the left boundary. The right boundary has homogeneous Neumann boundary conditions, representing null flux on that side.

In Figure~\ref{fig:domain:1d} we illustrate the setup for these tests. We represent in pink the pristine marble object, in cyan the outer air with $SO_2$ at concentration $s_{\text{ext}}$. The computational domain corresponds to the black thick line and the Dirichlet boundary condition is imposed on the left, where the stone is in contact with air. The right boundary is fictitious as the stone will continue past it and therefore we apply the homogeneous Neumann condition there.

\begin{figure}
\begin{center}
 \includegraphics[width=0.4\textwidth]{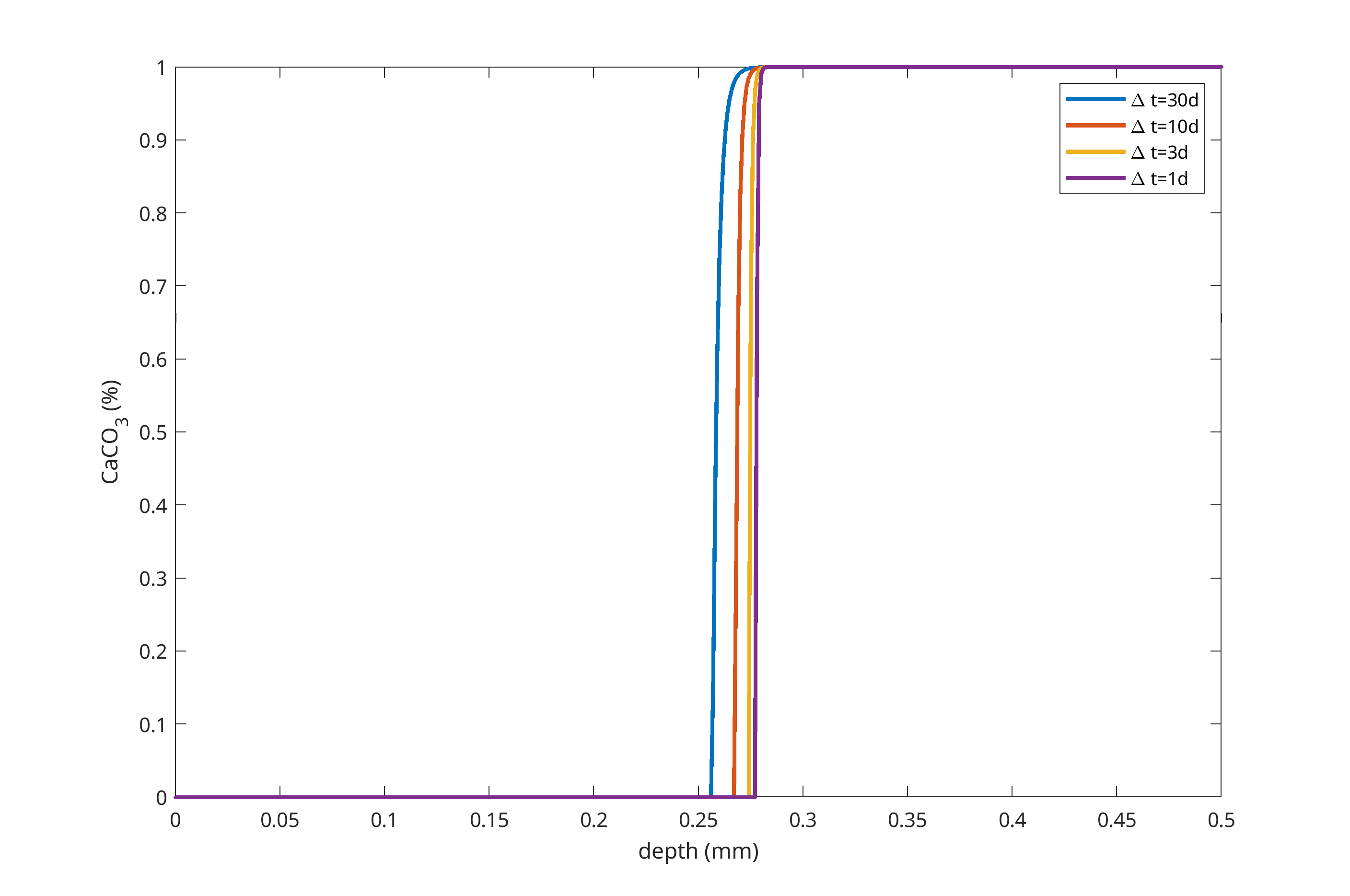}
 \includegraphics[width=0.4\textwidth]{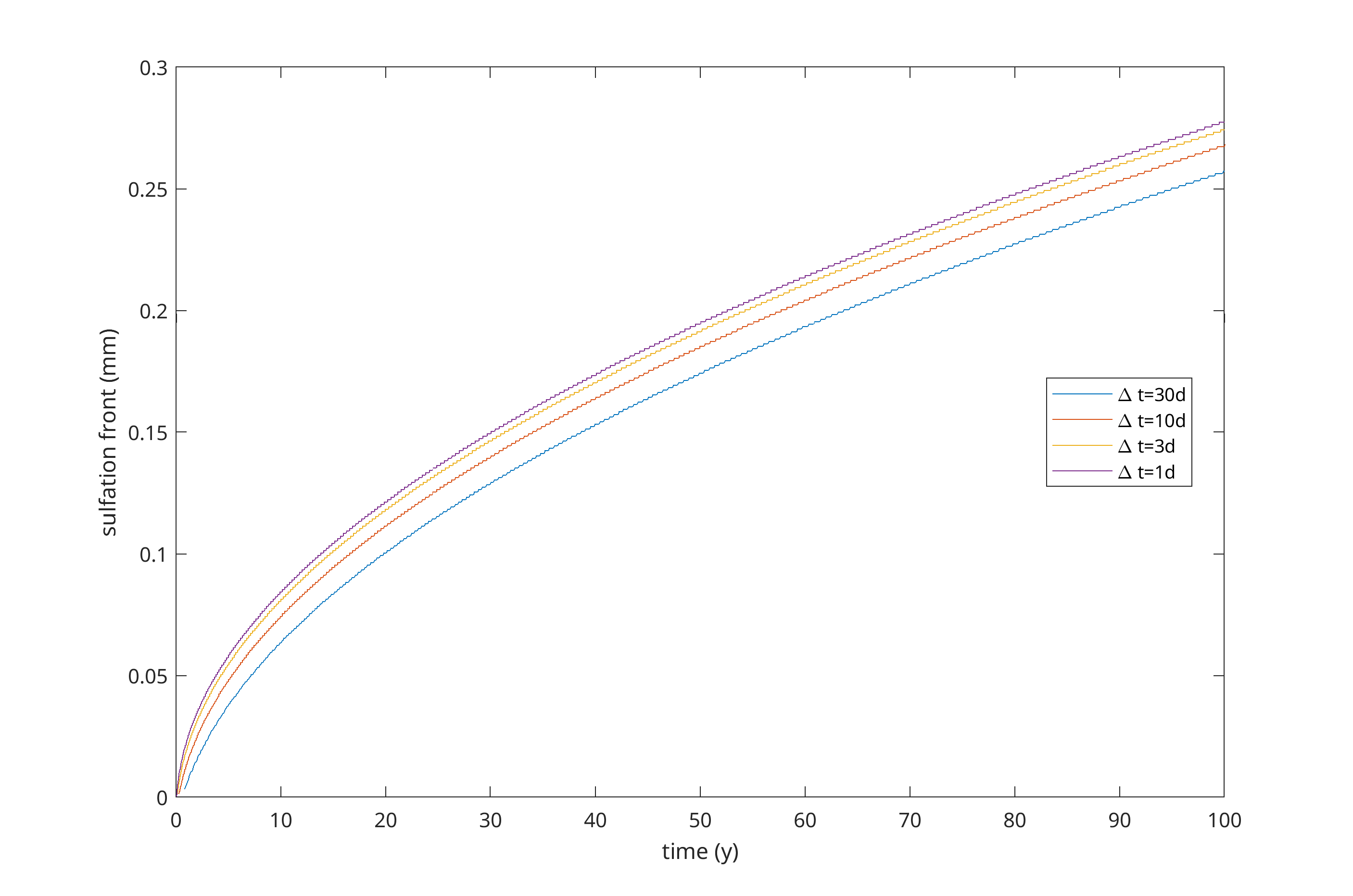}
\end{center}
\caption{Role of the discretization parameter $\Delta t$ on the front position and thickness.}
\label{fig:1d:dt}
\end{figure}

We start in Figure~\ref{fig:1d:dt}, where we present a simulation for $100$ years, computed in timesteps of length $30, 10, 3, 1$ day. The left plot depicts the $CaCO_3$ concentration with respect to the depth in the marble sample, illustrating the computed position and thickness of the sulfation front.
In the picture the ``sulfation front'' is the very small transition region across which the concentration of $CaCO_3$ drops suddenly from $1$ to $0$. Its thickness is very small (left panel) and its
position in time is depicted in the right panel, showing the typical square-root law proved in \cite{AFNT:07:darcymodel}. Except for the largest timestep choice, all the other results are very close to each other.

\begin{figure}
\begin{center}
 \includegraphics[width=0.4\textwidth]{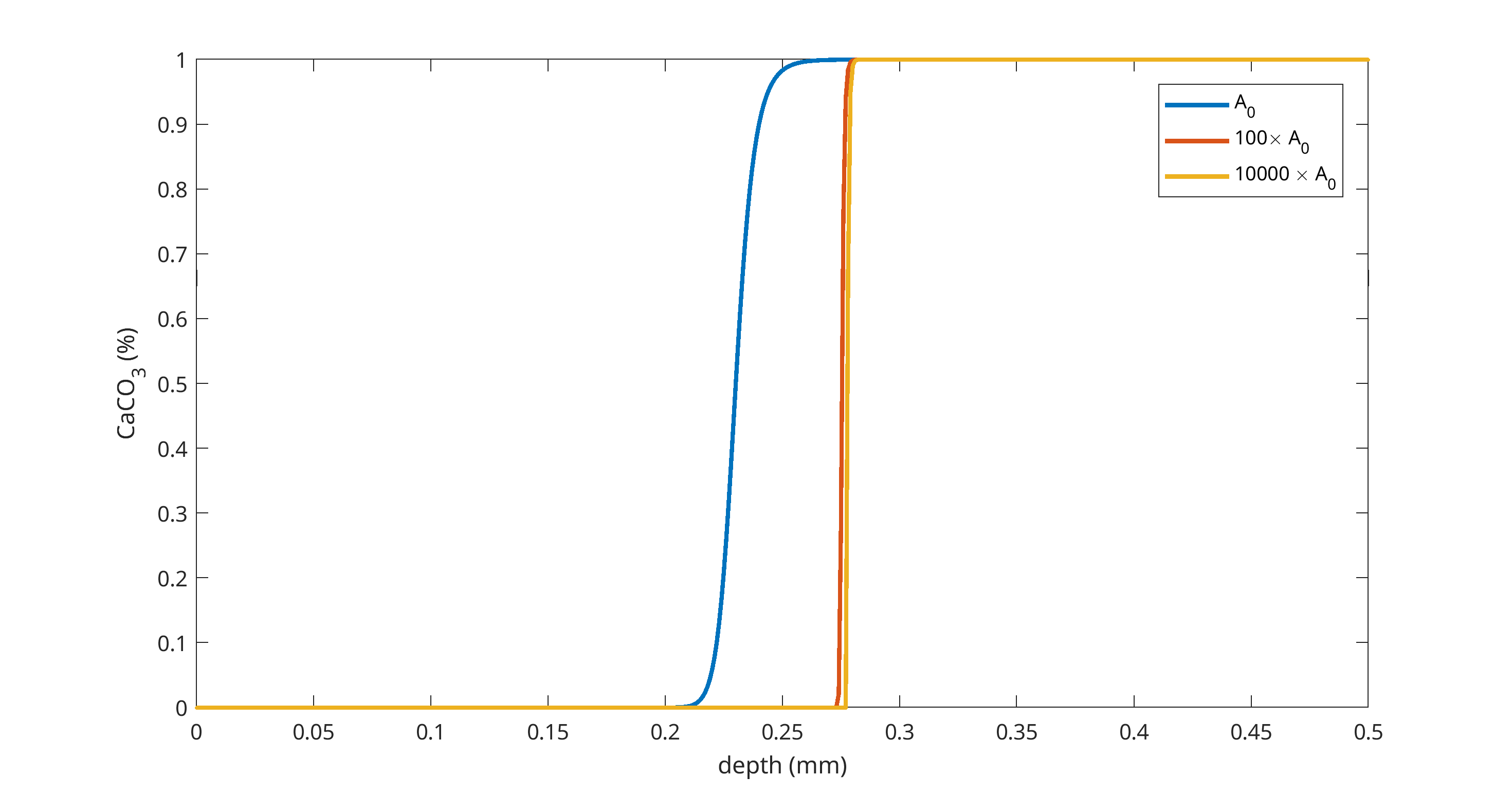}
 \includegraphics[width=0.4\textwidth]{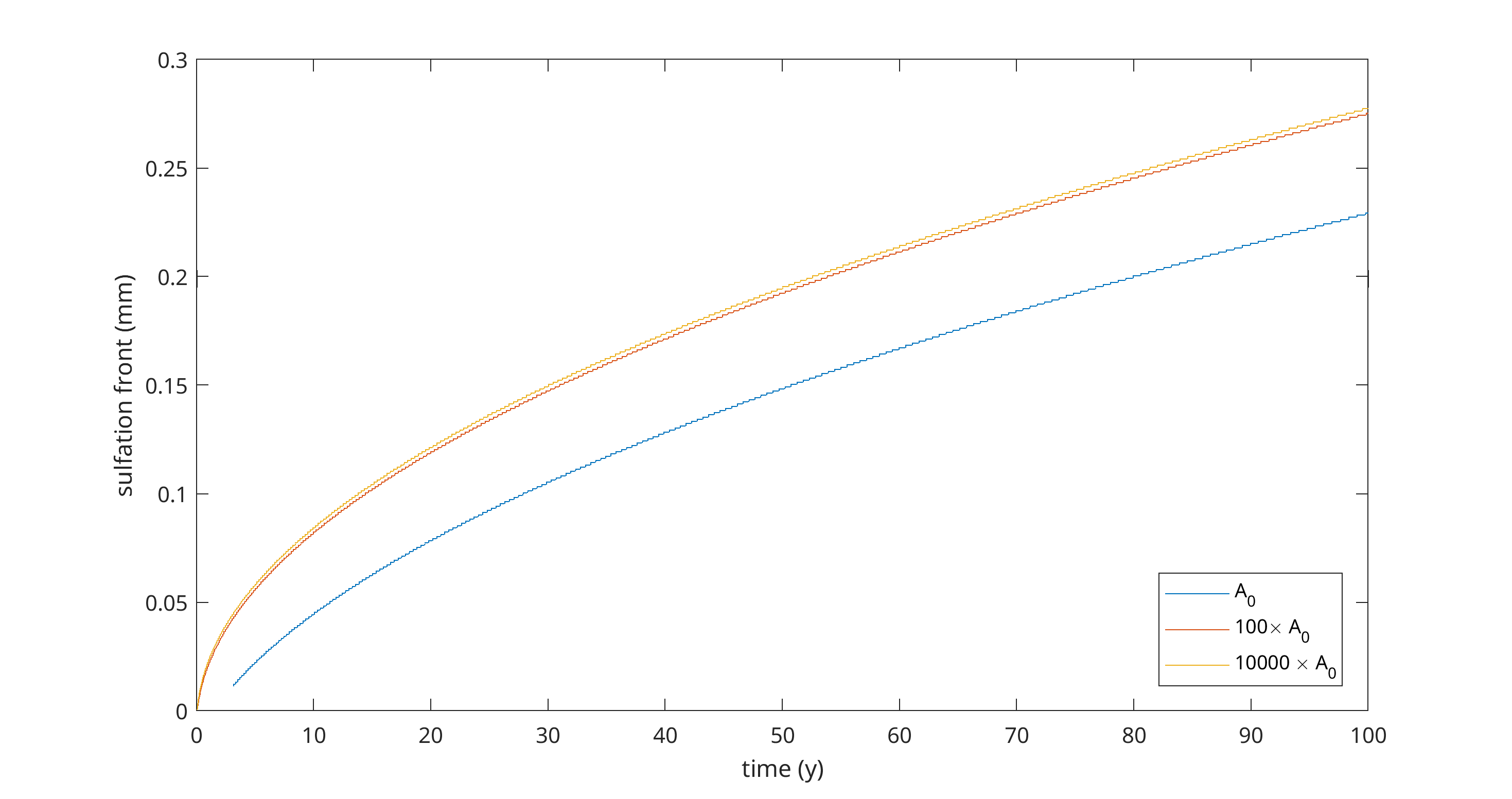}
\end{center}
\caption{Role of the reaction parameter $d$ on the front position and thickness.}
\label{fig:1d:A}
\end{figure}

Next we analyze the role of the reaction parameter $a$. In the real case it depends on the temperature and on the availability of water (vapour) at the reaction site. In Figure~\ref{fig:1d:A} we compare the simulations for $100$ years with different reaction rates. As it is clear from the plots, the reaction rate influences the thickness of the sulfation front.
For comparison, see for example the SEM image of a this slice of a reacting front in \cite{CFN:08:swelling}, where it can be appreciated that its thickness is just a few microns.
At the value $a=a_0=10^{-3}$, the sulfation front is $50\mu m$ thick and located at $220\mu m$ below the surface. However, from $a=10^{-1}$ onward, the front is very thin (few microns) and its position and speed does not change any more. This is the fast-reaction regime thus starts at this value and we will use $a=10^1$ in subsequent tests.
\todo{GB: una linea piu' spessa e un font un pò piu' grande potrebbero aiutare nella lettura dei risultati.}
Note that, once the fast reaction limit is reached, the penetration speed of the reaction front does not depend significantly on $a$.

\begin{figure}
\begin{center}
 \includegraphics[width=0.4\textwidth]{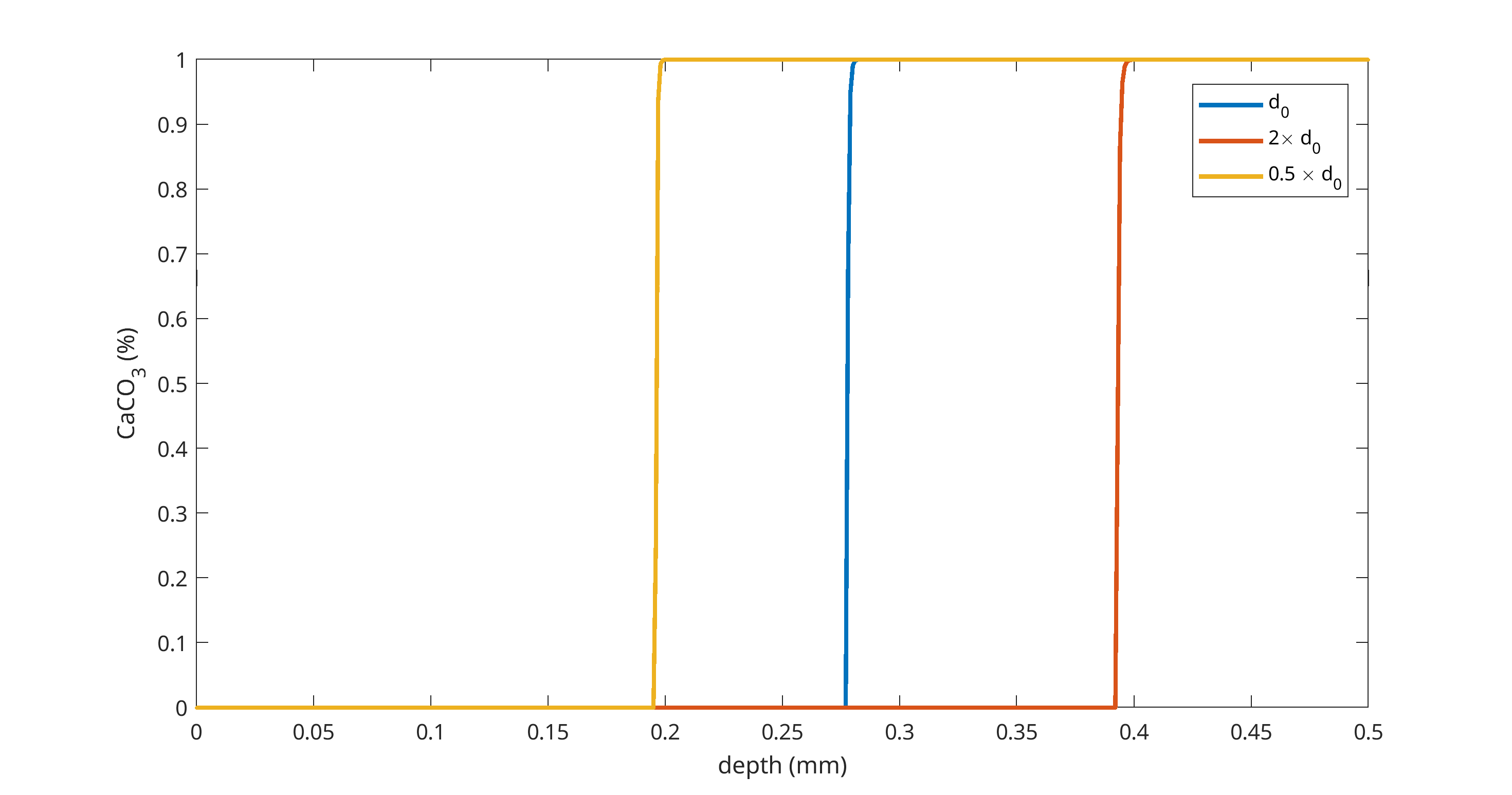}
 \includegraphics[width=0.4\textwidth]{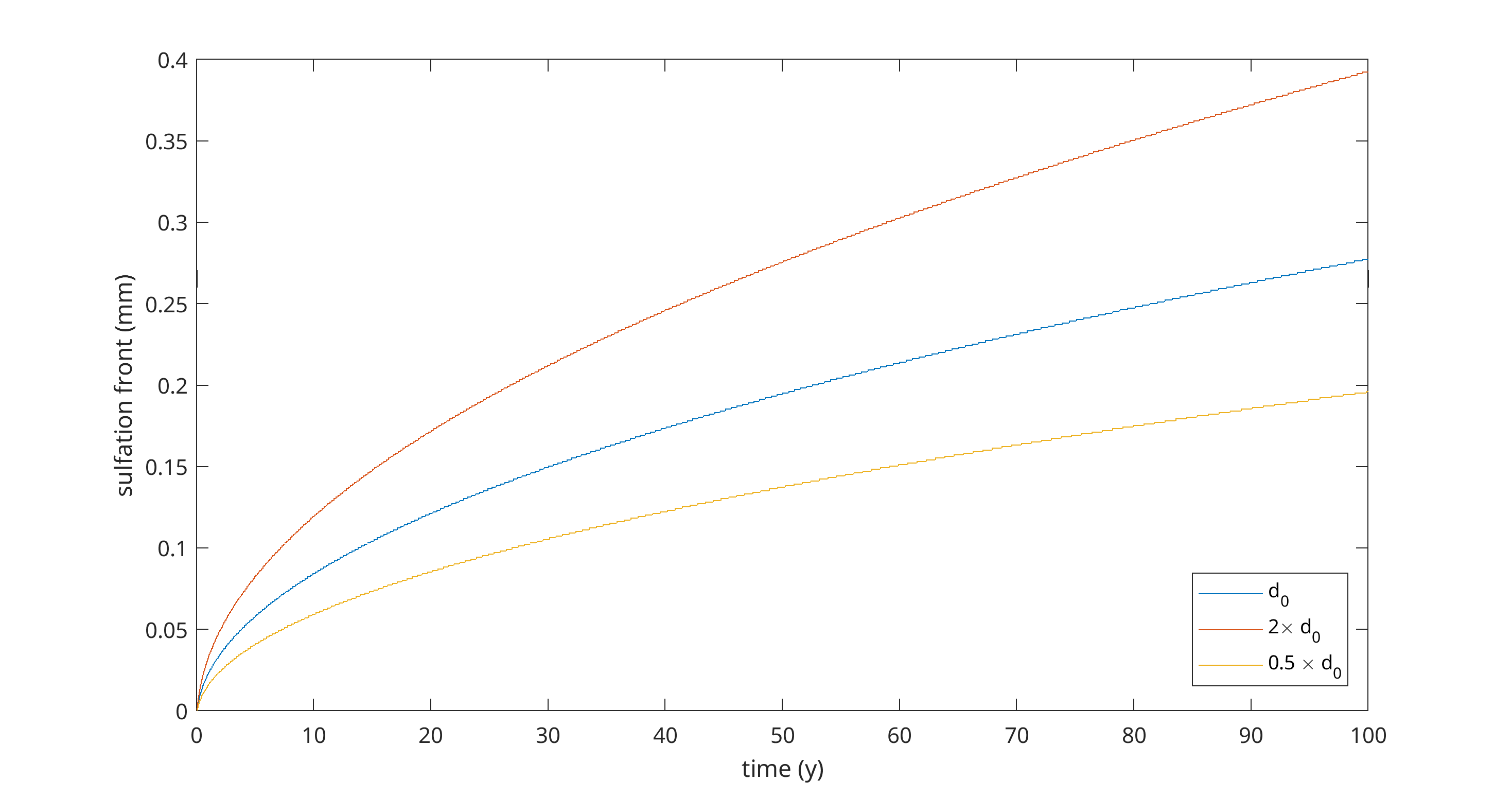}
\end{center}
\caption{Role of the diffusion parameter $d$ on the front position and thickness.}
\label{fig:1d:D}
\end{figure}

On the other hand, the parameter $d$ has a strong influence on the sulfation speed. Figure~\ref{fig:1d:D} shows three simulations for $d=d_0=10^{-10}$ and for a two-fold increased and decreased values of the diffusion coefficients. The very strong dependence of the phenomenon on this parameter is clear from the pictures.

\subsection{Sample 3D simulations}\label{ssec:3d}
In this section we consider the marble altar at Porta Marina spas, in the
area of the excavations of the Archaeological Park of Ostia Antica (see Fig.~\ref{fig:ostia}). A point cloud has been acquired by photogrammetry as described in Sec.~\ref{sec:pcloud}) and the corresponding levelset computed as reported in Sec.~\ref{sec:levelset}. The resolution of the point cloud is approximately $1mm$ and the generated levelset as about half that resolution. We ran our three-dimensional simulations on grids of size $0.5mm$, concentrating on small portions of the altar.

\paragraph{Damaged corner} We first consider a small portion of the altar, at the damaged corner in the upper part of the object.

\begin{figure}
\centering
    \includegraphics[width=0.31\linewidth]{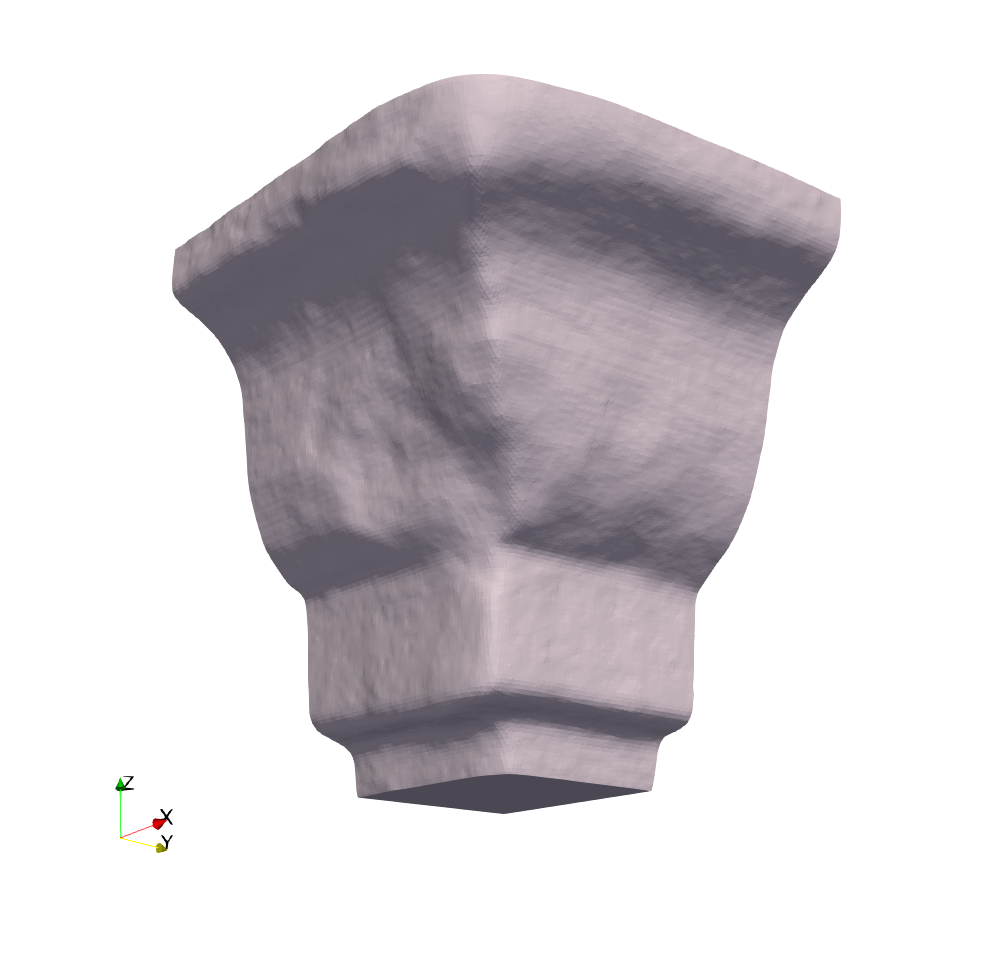}
    \includegraphics[width=0.31\linewidth]{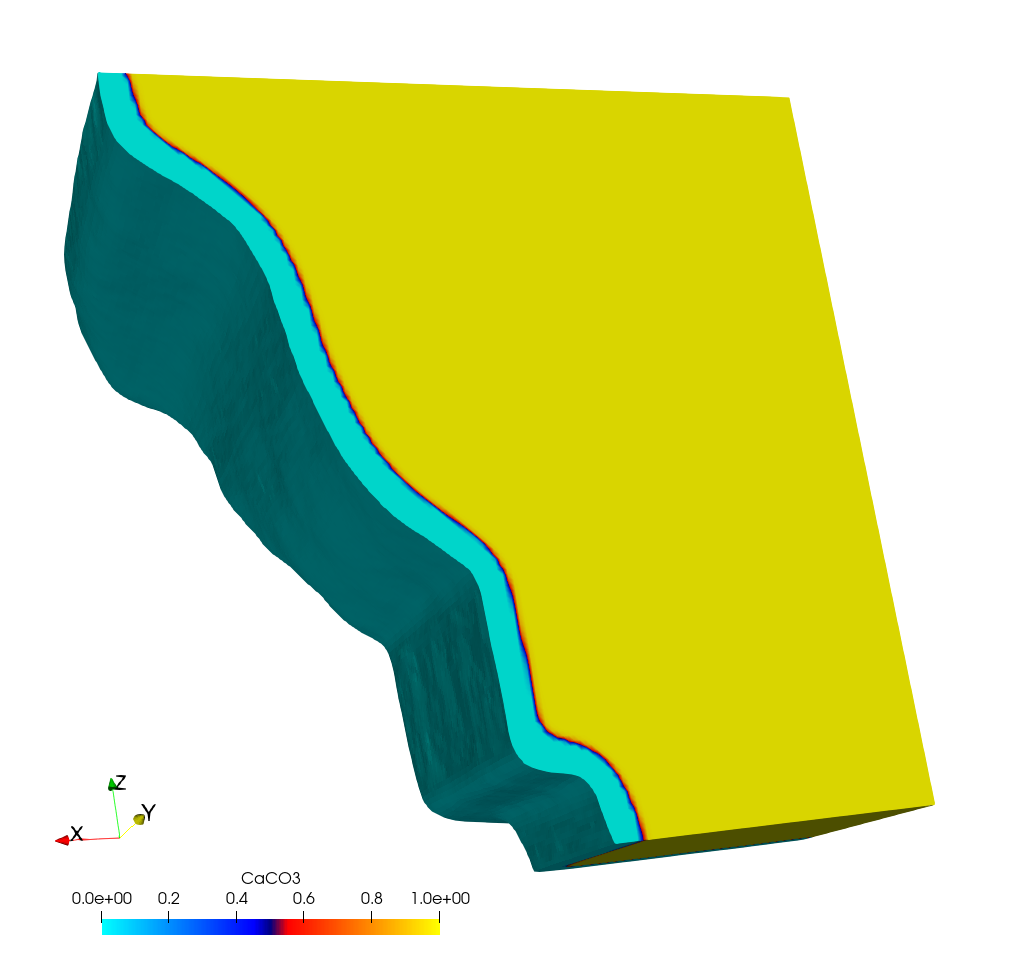}
    \includegraphics[width=0.31\linewidth]{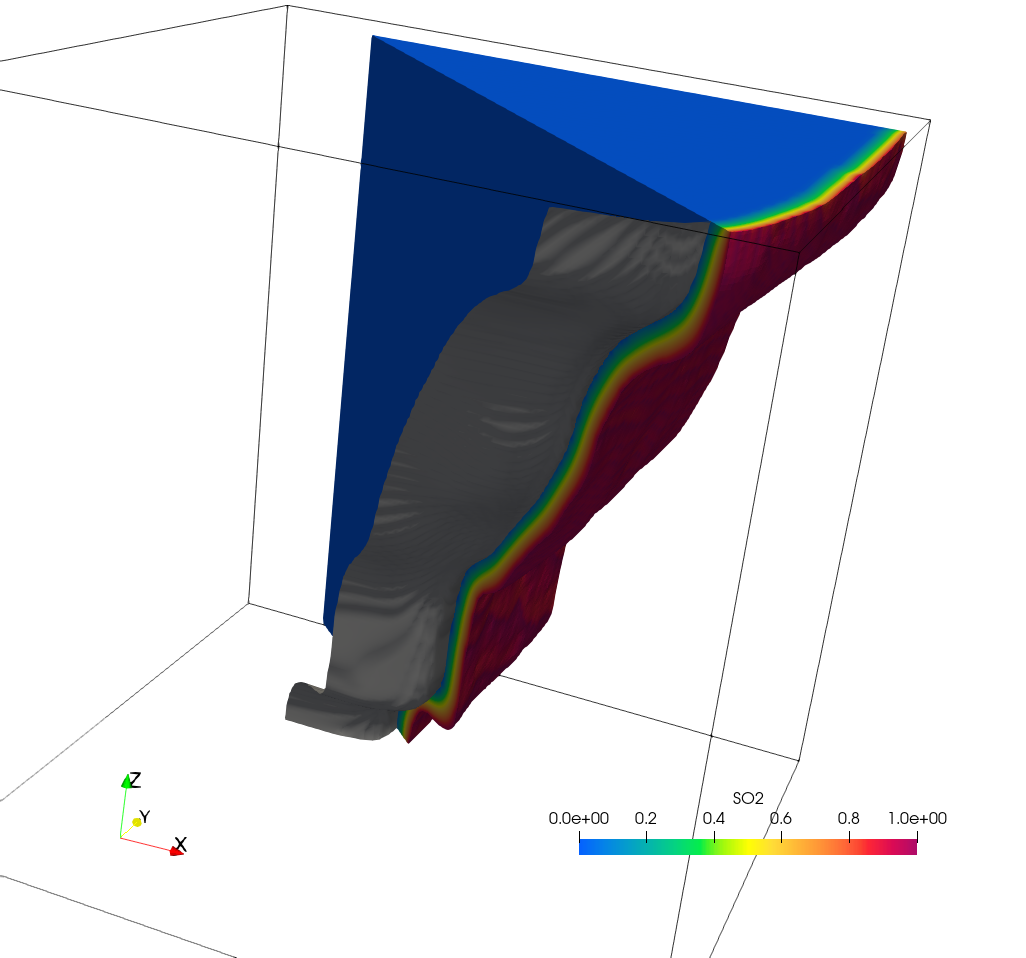}

\caption{Computation on the damaged upper corner. Left: levelset representing the outer surface of the domain. Center: $CaCO_3$ content in a section of the 3D domain, showing the thin sulfation front (blue-red transition). Right: $SO_2$ content in a section of the 3D domain (colors), and the isosurface $c=0.5$ (gray) representing the sulfation front.}
\label{fig:corner}
\end{figure}

This computation was performed on a grid of $3$ million points (precisely of size $156\times131\times150$) extracted from the levelset representing the corresponding part of the altar.\todo{qui abbiamo usato le half, non so se da lato alfora o da lato scritte} The resolution used in the PDE model was the same of the levelset, i.e. about $5\mathrm{mm}$.

In Figure~\ref{fig:corner} we present the output of a simulation. On the left we show the isosurface $\phi=0$ of the levelset, representing the outer surface of the object and of the computational domain. In the centre panel we plot the $CaCO_3$ residual content across a section cut diagonally near the corner of the domain. The cyan colour represents almost pure gypsum (i.e. fully degraded material), while the yellow colour represents pure marble (i.e. pristine material); the blue-red transition corresponds to $c\approx0.5$ and thus depicts the sulfation front at the time the snapshot was taken. It can be observed that the sulfation front is very thin. Also, it is noticeable how the thickness of the gypsum crust is influenced from the geometry of the outer surface: the crust is much thicker in convex areas of high curvature and thinner in flat or concave areas. The right panel depicts in colours the $SO_2$ concentration and in gray the sulfation front. It is clearly visible that the pollutant gas penetrates in the porous medium until the sulfation front and its presence is negligible below it, which is compatible with the fast reaction limit.

\paragraph{Inscription} Next we consider a portion of the face of the altar with the inscriptions. In this simulations we have extracted a $270\times65\times75$ subsample from the full levelset of this altar side, corresponding to the top left portion of the face with the inscriptions. The computational grid is composed of $1.3$ million points.

\begin{figure}
    \centering
    \includegraphics[width=.9\linewidth]{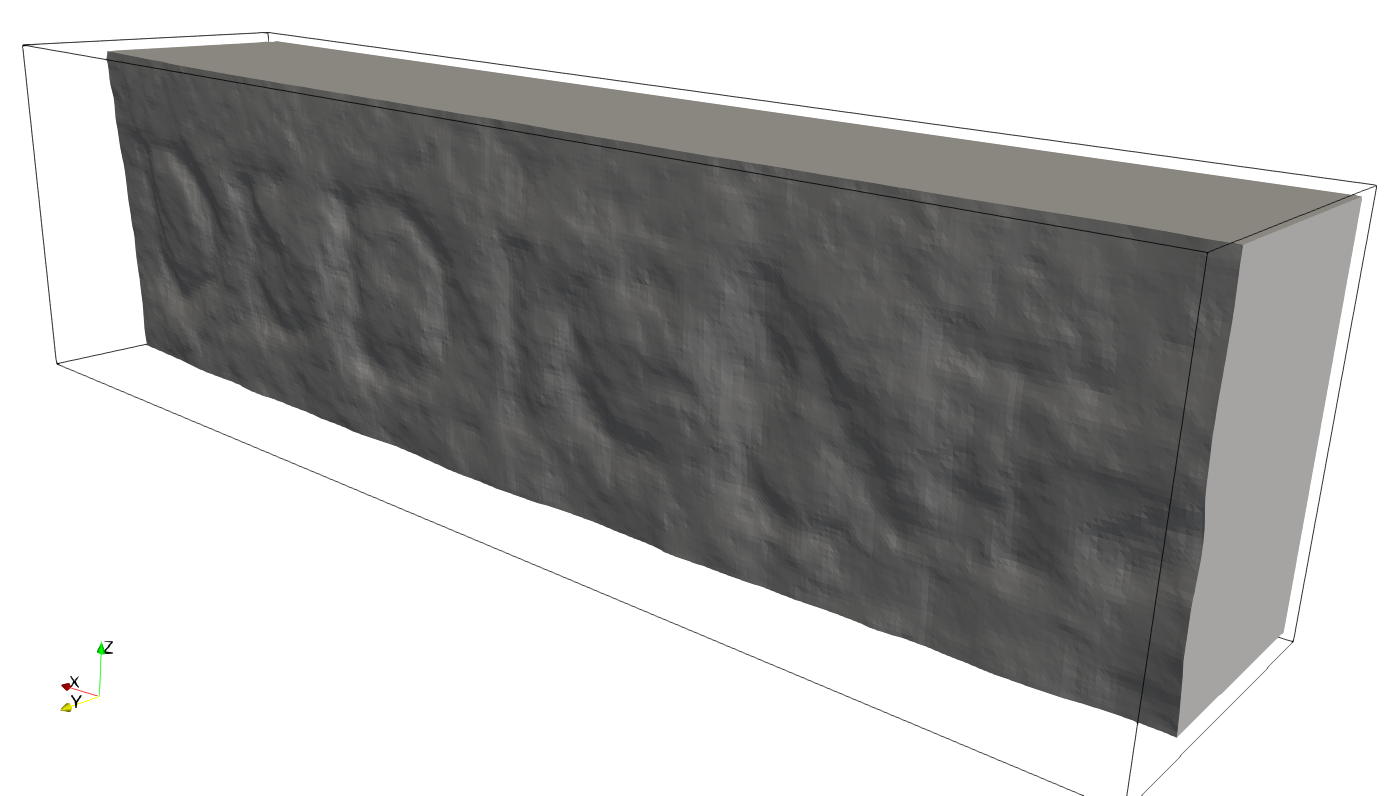}
    \\
    \includegraphics[width=.9\linewidth]{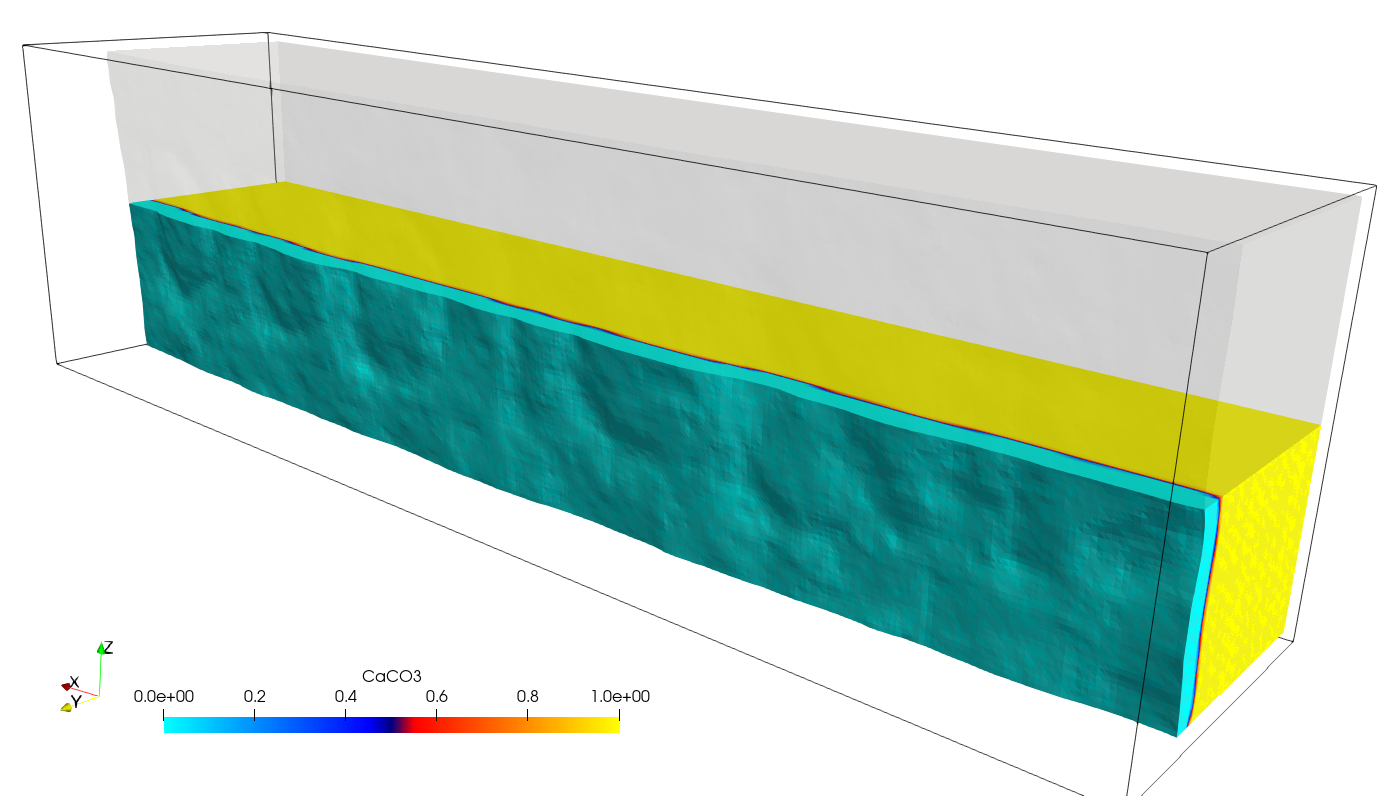}
\caption{Computation on the face with the inscriptions. Top: levelset representing the outer surface of the domain. Bottom: $CaCO_3$ content in a section of the 3D domain, showing the thin sulfation front (blue-red transition).}
    \label{fig:scritta}
\end{figure}

We present a snapshot of the simulation in Figure~\ref{fig:scritta}. In the top panel, we depict the Cartesian embedding domain (black edges) and the real computational domain defined by the levelset (gray surface). The bottom panel presents a section of the simulated domain, showing the internal concentration of $CaCO_3$. It is noticeable that the sulfation front, at the time of the snapshot, has overtaken the depth of the inscriptions and is almost completely flat; in this situation, a removal or dissolution by atmospheric agents of the gypsum crust would cause an almost complete loss of the inscriptions.

\paragraph{Face decoration} Here we consider a portion of the face of the altar with the circular decoration.
In this simulation we have extracted a $202\times65\times201$ subsample from the full levelset of this altar side, corresponding a central portion of the face with the inscriptions. The computational grid is composed of $2.6$ million points.
The simulated domain comprises a little more than a quarter of the circular decoration, from the rim (with a damaged portion in its bottom part) to the central ``button''.

\begin{figure}
    \centering
    \includegraphics[width=0.48\linewidth]{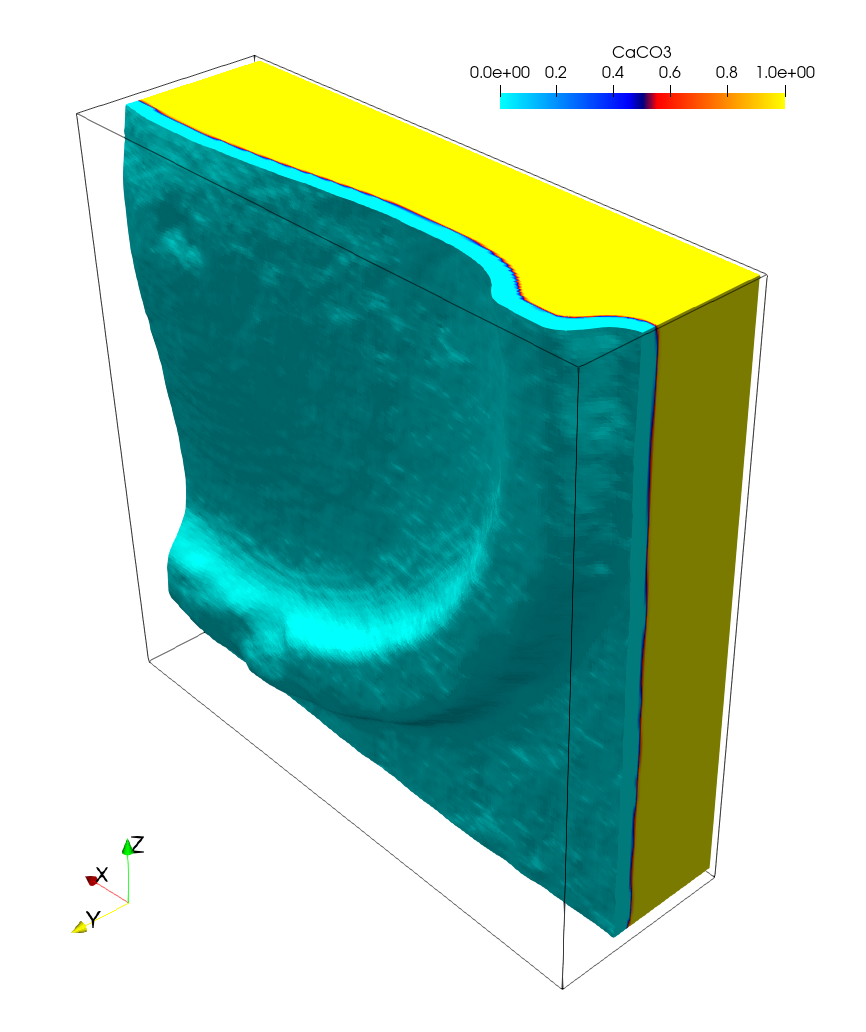}
    \hfill
    \includegraphics[width=0.48\linewidth]{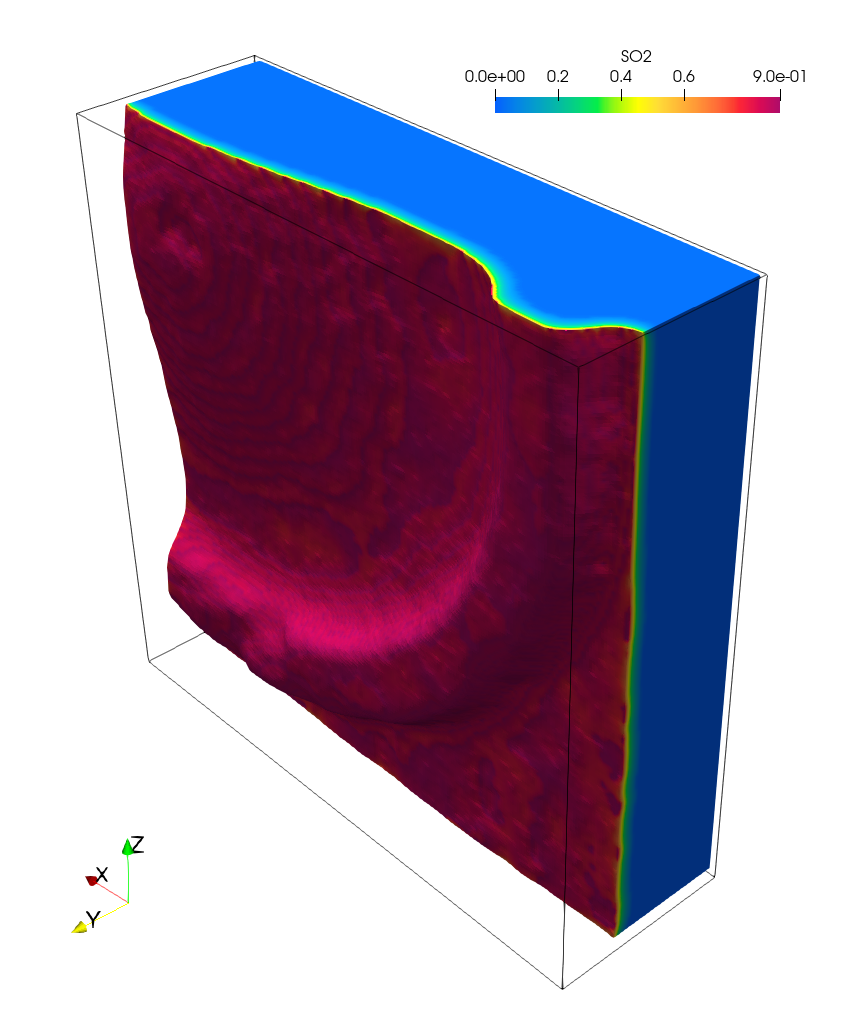}
    \caption{Final concentration of $CaCO_3$ and of $SO_2$ in the face decoration sample.}
    \label{fig:tondo:final}
\end{figure}

In Figure~\ref{fig:tondo:final} we depict the marble concentration (left panel) and the gas concentration (right panel) at the end of the simulation. The blue-red transition in the $CaCO_3$ content corresponds to a $50\%$ damaged material and represents the current sulfation front.

\begin{figure}
    \centering
    \begin{tabular}{cc}
      \includegraphics[width=0.48\linewidth]{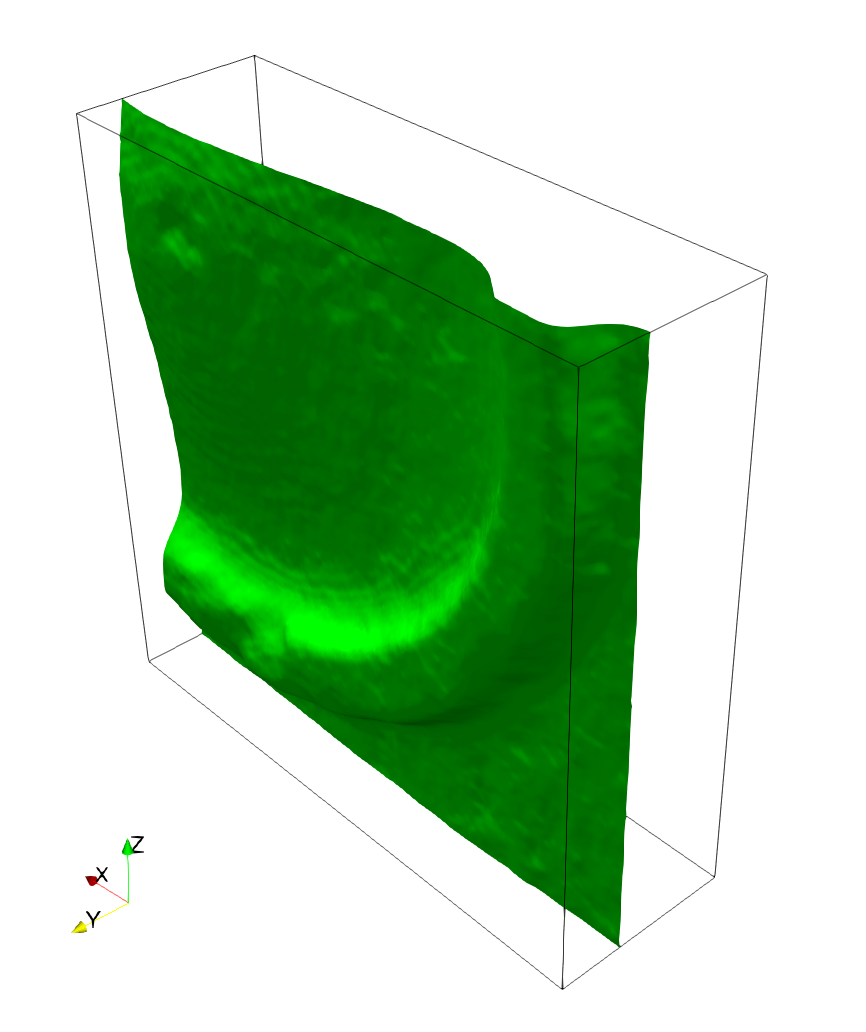}
      &
      \includegraphics[width=0.48\linewidth]{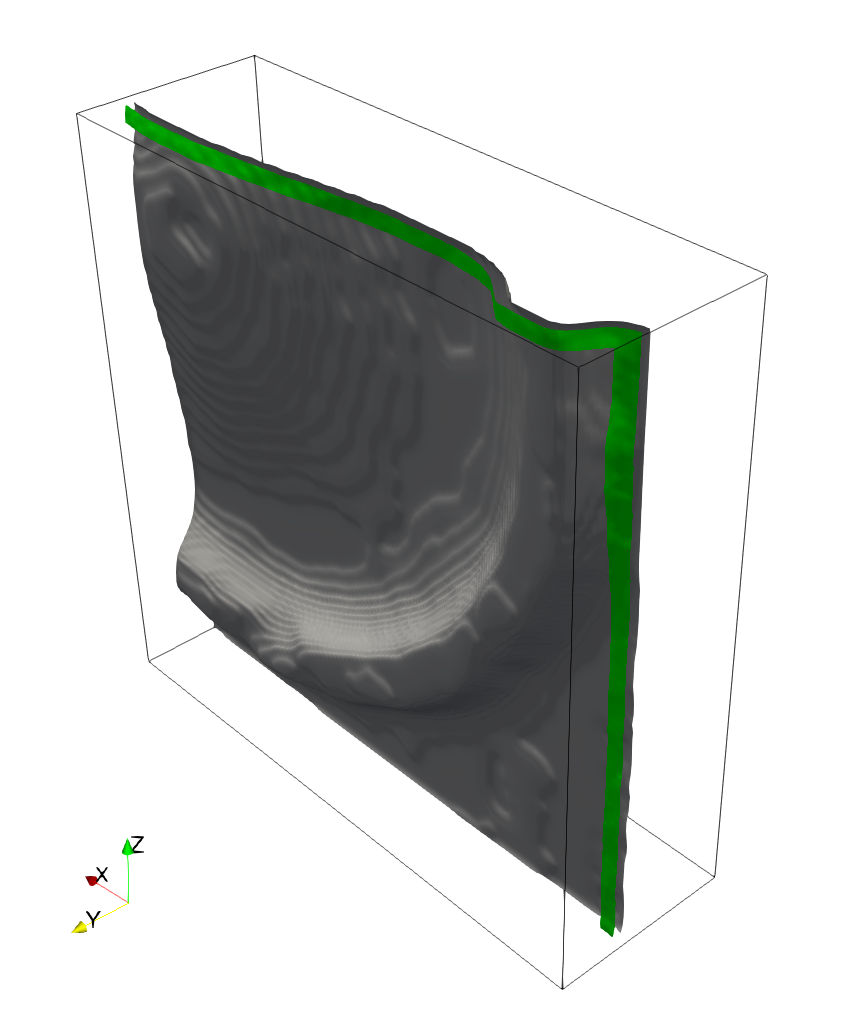}
      \\
      \includegraphics[width=0.48\linewidth]{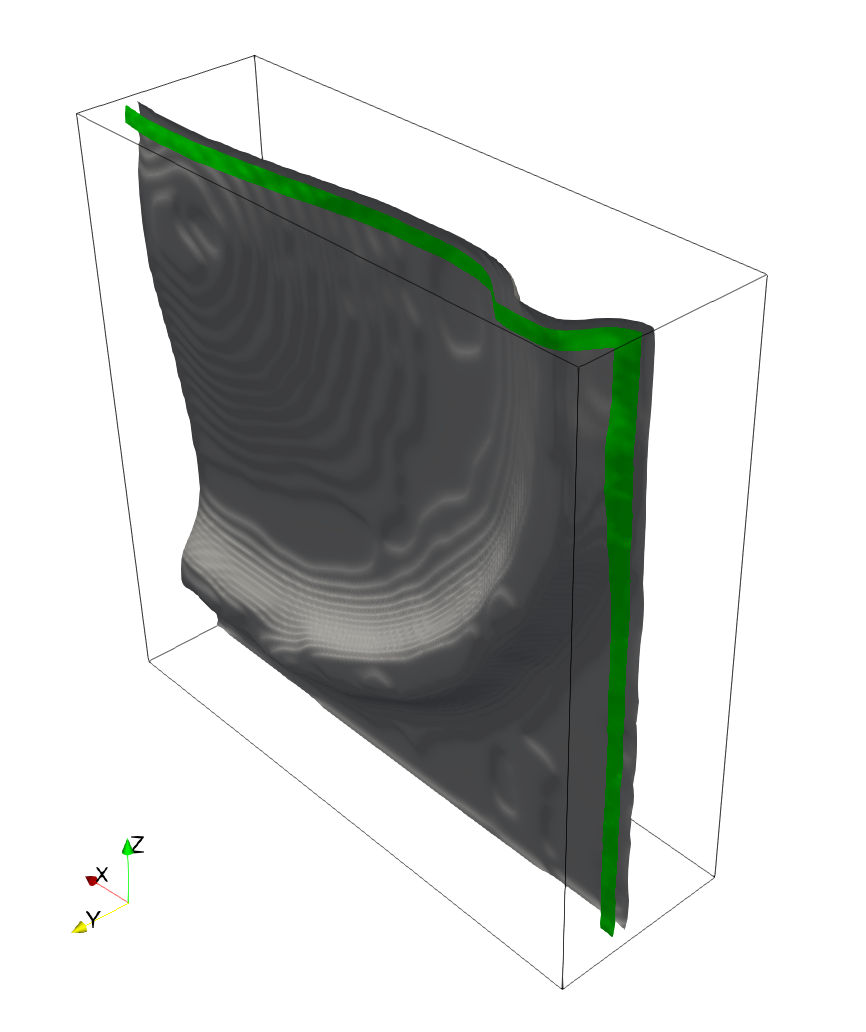}
      &
      \includegraphics[width=0.48\linewidth]{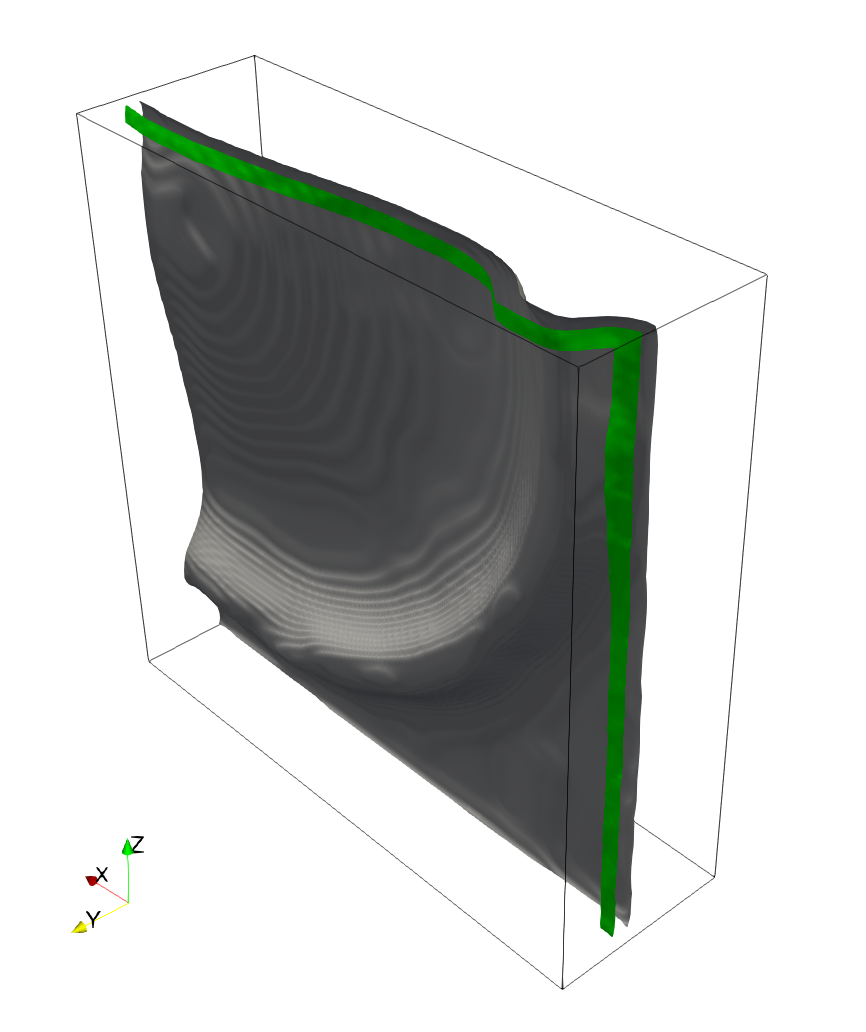}
    \end{tabular}

    \caption{Evolution in time of the sulfation front in the face decoration sample.  From top-left (initial) to bottom-right (final), the position and shape of the sulfation front is depicted in gray, while the green surface represents the initial outer surface.}
    \label{fig:tondo:evolution}
\end{figure}

The evolution of the sulfation front is illustrated in Figure~\ref{fig:tondo:evolution}. The top-left panel shows in green the original surface of the monument. The other three panels, which should be read from top-left to bottom-right, depict $c=0.5$ isosurface at increasing time during the simulation. This isosurface represents the evolution in time of sulfation front, which starts from the outer surface and gradually penetrates deeper inside the object.
A small portion of the outer surface in green is visible on the top and right edges to help appreciating the penetration of the sulfation front with time.
The sulfation front represents the shape that the object would have if the gypsum crust were to be removed or dilavated at that instant in time. It is clear that the central ``button'' would be almost immediately lost and that all sharp features would be rapidly smeared, leading to a severe loss in the readability of the decoration.

\begin{figure}
    \centering
    \includegraphics[width=0.75\linewidth]{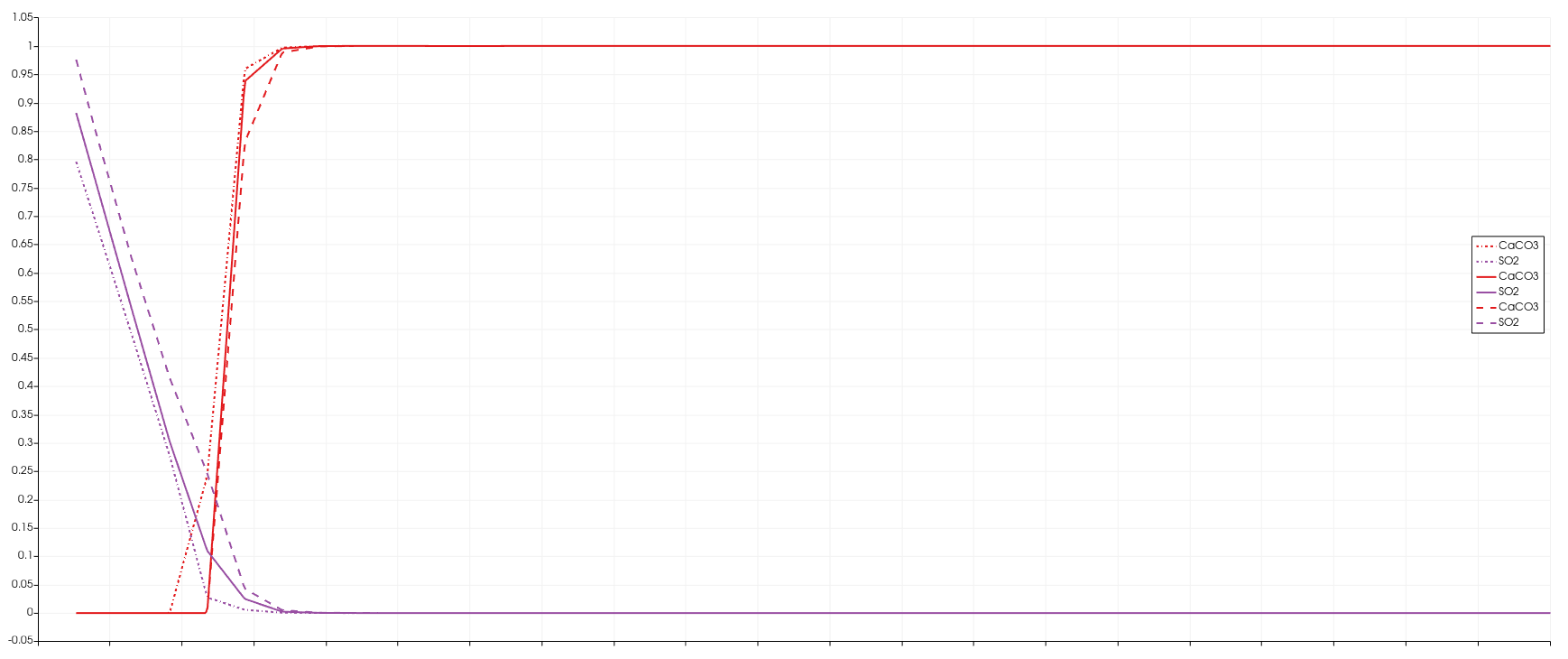}

    \includegraphics[width=0.75\linewidth]{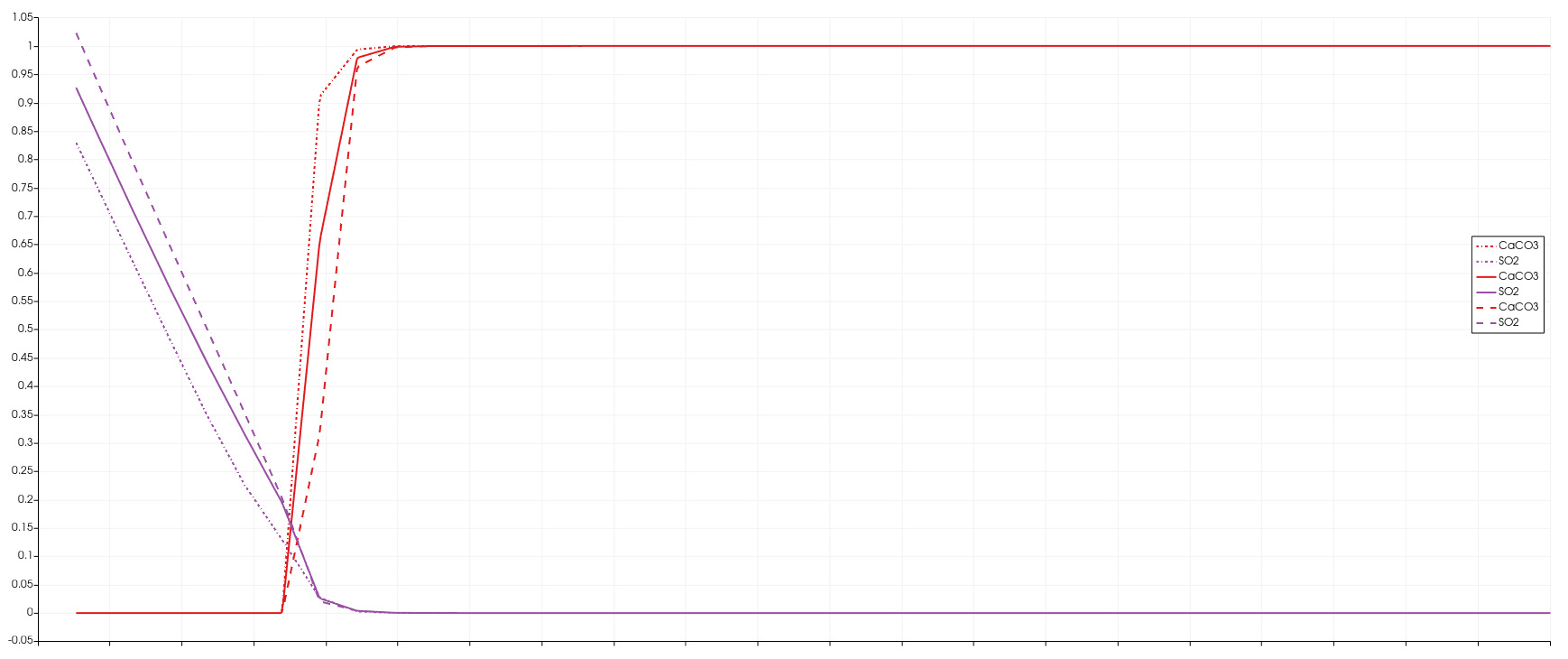}

    \includegraphics[width=0.75\linewidth]{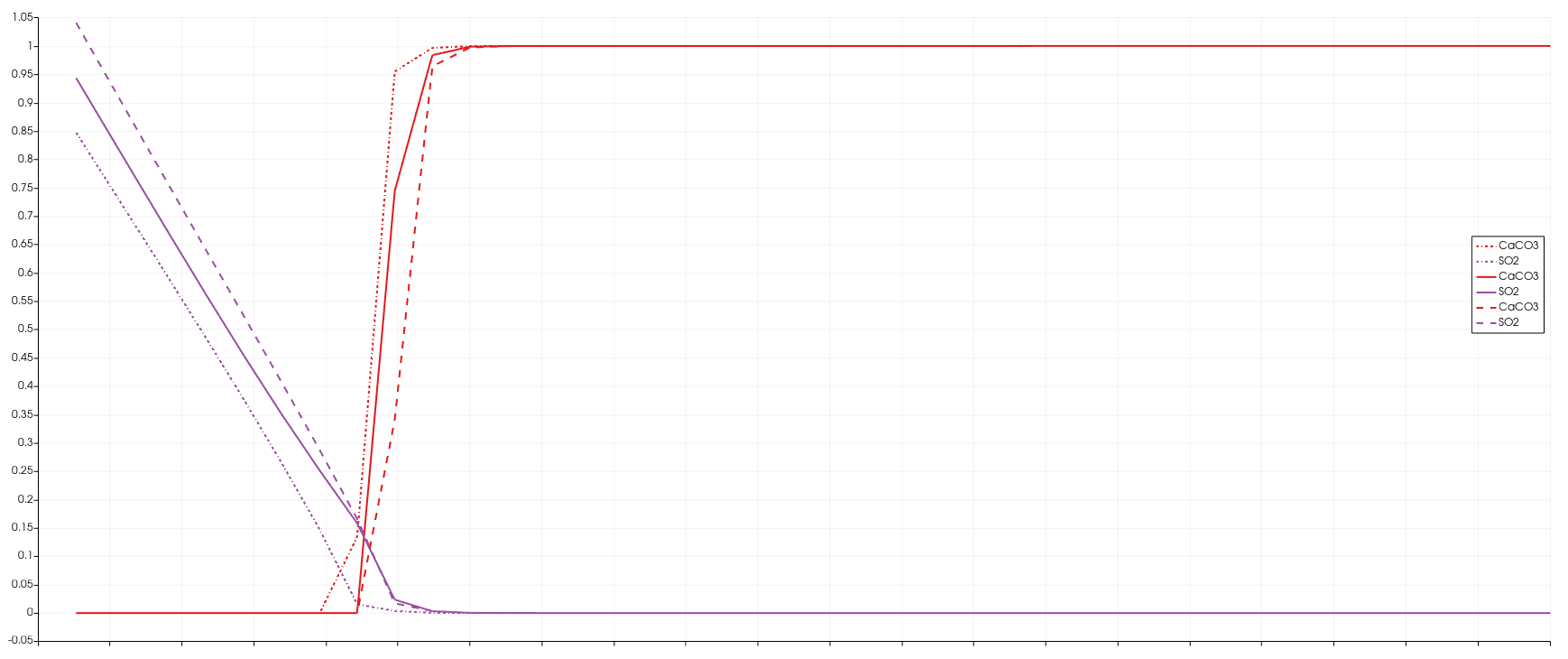}

    \caption{Plot of $CaCO_3$ and $SO_2$ concentrations along a line perpendicular to the outer surface of the rim of the circular decoration with the ``standard'' model parameters and with an external concentration of $SO_2$ reduced by $10\%$ (dash-dot lines) or increased by $10\%$ (dashed lines). From top to bottom are depicted the 2nd, 3rd and 4th time snapshots represented in Figure~\ref{fig:tondo:evolution}}.
    \label{fig:tondo:compare}
\end{figure}
In Figure~\ref{fig:tondo:compare} we depict the $CACO_3$ and the $SO_2$ concentration along a line perpendicular to the outer surface of the rim of the circular decoration, showing the position of the sulfation front (red lines) and the pollutant concentration inside the material (purple lines). From top to bottom, the panels refer to simulation times of the 2nd, 3rd and 4th time snapshots represented in Figure~\ref{fig:tondo:evolution}.
The simulation has been repeated with an increased (dashed lines) and decreased (dash-dot lines) external $SO_2$ concentration, in order to illustrate the effect of increasing/decreasing the amount of environmental pollution by $10\%$.

\paragraph{Computational times}
Finally, in Table~\ref{tab:times}, we report some data on the computational times of the simulation algorithm. For each of the three test cases reported in this section, we list the size of the background Cartesian grid, the number of preconditioned conjugate gradient iterations, the average time employed by the scheme to compute a timestep and the time difference between the cost of the first timestep with respect to the average. This latter is essentially dominated by the setup of the AMG preconditioner.

We observe that the number of preconditioned iterations of the main solver is not affected by the grid size and that the setup phase of the preconditioner has a moderate extra cost, corresponding to roughly $6$ timesteps. The number of seconds employed for the computation of each timestep is proportional to the grid size, indicating once more the optimality of the solvers employed.

The above tests were performed using $480$ cpus ($40$ cpus per $12$ fat nodes) on the Galileo100 cluster\footnote{https://www.hpc.cineca.it/systems/hardware/galileo100/} at CINECA. Each node is equipped with $2$ CPU Intel CascadeLake 8260,  with 24 cores each, 2.4 GHz, 384GB RAM and 3TB Intel Optane and the nodes are interconnected by an Infiniband network. The code was developed on the base of the PETSc 3.21.4 libraries \cite{petsc-web-page}.

\begin{table}
\centering

\begin{tabular}{lrrrr}
\toprule
case & grid size & CG iter & cpu/step (sec) & AMG setup (sec)\\
\midrule
corner & $3.06\times10^6$ & $3\div4$ & 4.16 & 23.0\\
script & $1.31\times10^6$ & $3\div4$ & 1.64 & 11.3\\
decoration & $2.63\times10^6$ & $3\div4$ & 3.35 & 17.2\\
\bottomrule
\end{tabular}

\caption{Computation time data on $480$ cores.}
\label{tab:times}

\end{table}

\section{Concluding remarks and future perspectives}
\label{sec:concl}

We have proposed a complete framework to perform ``in silico'' simulations and experiments about the phenomenon of stone degradation in cultural heritage applications. The method involves (a) the acquisition of a point cloud describing the exact shape of the object, (b) the construction of a 3D model of the object by computing a levelset function and (c) the numerical simulations by a PDE-based model via ghost-point finite-difference methods.

In this paper we have chosen photogrammetry for point (a), which is easily applicable to acquire in situ the shape of large objects without the need to transport and install specialized equipment. Of course, for smaller or indoor objects, laser scanning is a viable alternative.

For points (b) and (c) we have used well-tested numerical methods on uniform Cartesian grids. However, the levelset evolution method of (b) is now available also on locally adaptive octree grids \cite{PreSe:26:adaptive} and it would be interesting to extend also the numerical method to this setup, which should grant computational time advantages, since a fine grid is really needed only in the vicinity of a thin evolving front where the degradation is occurring. To this end, also finite element based methods like the ghost-FEM presented in \cite{Astuto2025} is an interesting alternative as it simplifies the treatment of boundary conditions.

For this paper we have chosen to apply a simple mathematical model of marble sulfation \cite{ADN:sulfation}, which is composed by a reaction-diffusion equation for the pollutant gas penetration into the pores of the material and by a pure reaction equation for the material degradation. Due to the stiffness of the reaction, we have adopted an operator splitting approach composed by an exponential evolution operator for the reaction terms and by an implicit ghost finite-difference approach for the parabolic diffusion terms. 

We have shown how to conduct with the proposed method ``in silico'' experiments on the phenomenon of marble sulfation, understanding the role of model parameters (\S\ref{ssec:sensitivity}) and the effect of local geometry and of varying the environmental conditions on the thickness of the degradation crust (\S\ref{ssec:3d}).
We point out that the same numerical approach could be exploited to perform ``in silico'' experiments with different or more complex mathematical models of interest in cultural heritage applications.
\todo[inline]{quali citiamo come esempi?

\textcolor{magenta}{GB: possiamo dire che si possono sviluppare modelli che combinano fenomeni di erosione e formazione di crosta ed anche altri fenomeni come la cristallizzazione o la formazione di biofilm? }
}

In this regards, the computational framework developed for this study will be included in the Stoneverse platform\footnote{\url{https://stoneverse.iac.cnr.it}}, see also \cite{stoneverse}, with the aim of sharing knowledge on new technologies and supporting  methodological developments within the Cultural Heritage scientific community.

\paragraph*{Acknowledgements}

This work has been funded by the Italian Ministry of University and the European Union, under the PRIN-PNRR project Prot.\ P20228HZWR.

G.~B., S.~P. and M.~S.\ are members of the Gruppo Nazionale Calcolo Scientifico-Istituto Nazionale di Alta Matematica (GNCS-INdAM).

M.~S. and S.~P. acknowledge the CINECA award under the ISCRA initiative, for the availability of high performance computing resources on Galileo~100 and for the support.

\bibliographystyle{abbrv}
\bibliography{sulfation.bib}
\end{document}